\documentclass{iopart}

\usepackage{graphicx}
\usepackage{psfrag}

\usepackage[colorlinks]{hyperref}

\begin{document}

\title[Laser cooling in the Penning trap]{Laser cooling in the Penning trap: an analytical model for cooling rates in the presence of an axializing field}

\author{R~J~Hendricks\footnote{Now at: Institute for Physics and Astronomy, University of Aarhus, DK-8000~\AA rhus~C., Denmark}, E~S~Phillips\footnote{Now at: Clarendon Laboratory, University of Oxford, Parks Road, Oxford OX1 3PU, UK}, D~M~Segal and R~C~Thompson}

\address{Blackett Laboratory, Imperial College, London SW7 2AZ, UK}

\ead{r.thompson@imperial.ac.uk}

\begin{abstract}
Ions stored in Penning traps may have useful applications in the field of quantum information processing.  There are, however, difficulties associated with the laser cooling of one of the radial motions of ions in these traps, namely the magnetron motion.  The application of a small radio-frequency quadrupolar electric potential resonant with the sum of the two radial motional frequencies has been shown to couple these motions and to lead to more efficient laser cooling.  We present an analytical model that enables us to determine laser cooling rates in the presence of such an `axializing' field.  It is found that this field leads to an averaging of the laser cooling rates for the two motions and hence improves the overall laser cooling efficiency.  The model also predicts shifts in the motional frequencies due to the axializing field that are in qualitative agreement with those measured in recent experiments.  It is possible to determine laser cooling rates experimentally by studying the phase response of the cooled ions to a near resonant excitation field.  Using the model developed in this paper, we study the expected phase response when an axializing field is present.
\end{abstract}


\maketitle


\section{Introduction}
\label{sec:introduction}

Quantum information processing (\textrm{QIP}) has been the subject of much research in recent years because of its potential ability to provide a polynomial or even exponential speedup over con\-ven\-tional computing when solving certain problems, perhaps most notably the factoring of large integers, the searching of unsorted lists and the simulation of quantum systems~\cite{Spiller05,N&C}.  A key requirement for any system that is to form the basis of a quantum computer is that it must be well isolated from its environment in order to prevent decoherence.  For this and other reasons, ions trapped in radiofrequency (\textrm{RF}) traps have proved to be an excellent system and many of the basic requirements for QIP have been demonstrated using them~\cite{NIST03:gate,Schmidt-Kaler03:CNOT,NIST05:6ionGHZ,Haffner05,NIST05:fieldinsensitivequbits,Haffner05:qubitlifetime}.  However, the large amplitude oscillating fields used for confinement in \textrm{RF} traps mean that ion heating rates can be rather large and can lead to a loss of coherence during gate operations~\cite{Michigan04:heatingrates,Michigan06:needletrapheating}.  With its purely static electric and magnetic fields, the Penning trap may therefore provide an attractive alternative system for research towards \textrm{QIP}.  The operation of a universal quantum computer based on quantum gates is extremely challenging since it requires very high fidelity and low decoherence.  On the other hand, direct quantum simulation (\textrm{DQS}) in which trapped ions are used to directly simulate other quantum systems may prove to be a more tractable experimental prospect.  An array of ions in a Penning trap has recently been proposed as an ideal system in which to perform \textrm{DQS} of a range of systems of interest to the condensed matter community~\cite{Cirac04a,Cirac04b,Cirac06}.  Work is also being carried out on the use of novel Penning trap designs for quantum information processing~\cite{Stahl05,IC06}.

By using a Penning trap located in a superconducting magnet it is also possible to reduce the influence of fluctuations in the ambient magnetic field, which currently limit the coherence times that are observed in many \textrm{RF} trap experiments.  In order to combat the effects of fluctuating magnetic fields there has been much recent interest in the use of first-order magnetic field free transitions as qubit transitions in \textrm{QIP} studies.  These are transitions between two internal electronic states whose Zeeman shift gradients are at some point equal, so that the transition frequency is (to first order) insensitive to changes in the magnetic field~\cite{NIST05:fieldinsensitivequbits}.  In many cases these transitions occur at rather large magnetic fields, making them well suited for use with ions trapped in a Penning trap~\cite{Bollinger91}.

The two key difficulties with working in the Penning trap are associated with the laser cooling of the ions.  Firstly, the large magnetic field used for trapping (typically a few tesla) leads to rather large Zeeman splittings and so the number of transitions that must be addressed to form a closed cooling cycle is greater than in a \textrm{RF} trap.  This increases the technical complexity of an experiment, but has not proved insurmountable~\cite{Wineland78,Koo04}.  The second difficulty is that the ions in a Penning trap have a rather more complex motion that includes an unstable component, called the magnetron motion.  Under normal circumstances laser cooling of this component of the motion is necessarily inefficient.  A technique called axialization has been developed and demonstrated to improve the efficiency of laser cooling in the Penning trap by coupling the two radial motions~\cite{Powell02,Powell03}.  An analytical model for axialization is presented in this paper.  This model exhibits a number of features that have been observed in recent experimental work on the axialization technique. 

We first review both ion motion in a Penning trap and a simple analytical model for laser cooling of the two radial motions.  We then proceed to introduce the effect of the coupling of these two motions into the model and derive damping rates in the presence of axialization.  Finally, we examine how an external excitation of one of the motions can reveal information on the damping rates and the coupling rate between the motions.


\section{Motion in the Penning trap}
\label{sec:penningmotion}

\begin{figure}
\begin{center}
\includegraphics[width=1.0\textwidth]{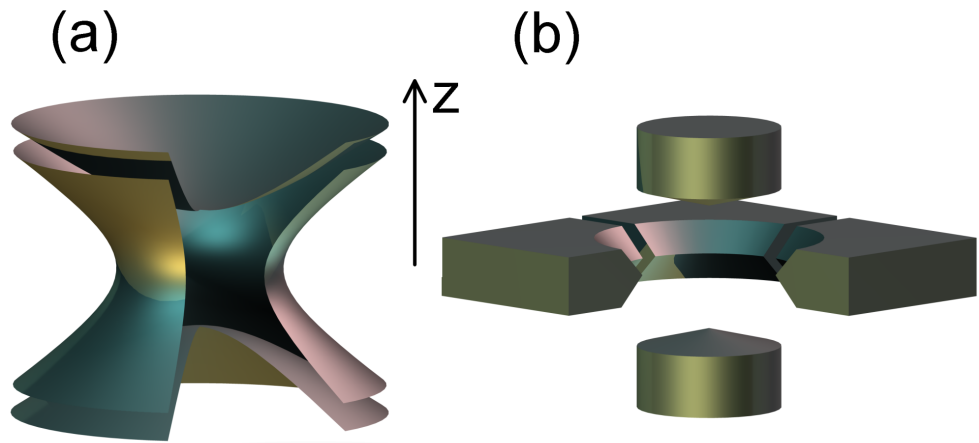}
\caption{\textbf{(a)} Electrode structure of an idealised Penning trap.  \textbf{(b)} Non-ideal Penning trap with a split ring electrode.  This is similar to a trap used in recent experiments.  Note that a section from the left image and one of the ring segments from the right image have been removed for clarity.
\label{fig:penningtrapstructure}}
\end{center}
\end{figure}

A detailed description of the Penning trap can be found in, for example,~\cite{Ghosh,Werth}.  In brief, an idealised Penning trap consists of electrodes whose inner surfaces are hyperbolae of revolution about the $z$ axis (figure~\ref{fig:penningtrapstructure}(a)).  For positively charged ions the two endcaps are given a positive bias with respect to the ring electrode.  This leads to a harmonic electric potential $\phi$ that confines in the $z$ direction, but results in an outward force in the radial plane.  Radial confinement is achieved by applying an additional static magnetic field of typically a few tesla along the axis of the trap.  This forces the ions into cyclotron-like loops.  In practice it is often convenient to use electrode shapes that differ somewhat from the ideal.  Furthermore, in order to apply the axialization technique described in this paper it is necessary for the ring electrode to be split into four radial segments (see figure~\ref{fig:penningtrapstructure}(b)).  This enables an additional small radially quadrupolar electric potential to be applied.

The force on an ion with charge $e$ due to the trapping field is given by the Lorentz force
\begin{equation}
\label{eq:Lorentz force}
\bi{F}=e\bi{v} \times \bi{B} - e \mathbf{\nabla} \phi,
\end{equation}			
where $\bi{B} = - B\hat{\bi{z}}$ and $\phi = U_0 \left(  2z^2 - x^2 - y^2 \right)  / \left(  2z_0^2 + r_0^2 \right) $.  In the $z$ direction this results in simple harmonic motion with a frequency 
\begin{equation}
\label{eq:axialfrequency}
\omega_{\mathrm{z}} = \sqrt{\frac{4eU_0}{m(2z_0^2 + r_0^2)}}.
\end{equation}
In the $x$ and $y$ directions (\ref{eq:Lorentz force}) leads to the coupled equations
\begin{eqnarray}
\label{eq:radialcoupledequations}
\ddot{x}  + \omega_{\mathrm{c}} \dot{y} - \frac{\omega_{\mathrm{z}}^2}{2} x & = & 0, \nonumber\\
\ddot{y} - \omega_{\mathrm{c}} \dot{x} - \frac{\omega_{\mathrm{z}}^2}{2} y & = & 0.
\end{eqnarray}
Here $\omega_{\mathrm{c}}$ is the true cyclotron frequency $eB/m$.  These equations can be solved by defining a new variable $u=x+iy$, so that
\begin{equation}
\label{eq:PenninguDE}
\ddot{u} - i \omega_{\mathrm{c}} \dot{u} - \frac{\omega_{\mathrm{z}}^2}{2} u = 0.
\end{equation}
Trying a solution corresponding to circular motion, $u=u_0 \mathrm{e}^{i \omega t}$, leads to
\begin{equation}
\label{eq:PenninguDEsolution}
u_0 \mathrm{e}^{i \omega t} \left( \omega^2 - \omega_{\mathrm{c}} \omega + \frac{\omega_{\mathrm{z}}^2}{2}\right)  = 0.
\end{equation}
This has roots
\begin{eqnarray}
\label{eq:PenninguDEroots}
\omega & = & \frac{\omega_{\mathrm{c}}}{2} \pm \sqrt{\frac{\omega_{\mathrm{c}}^2}{4} - \frac{\omega_{\mathrm{z}}^2}{2}} \nonumber \\
 & = & \frac{\omega_{\mathrm{c}}}{2} \pm \omega_1,
\end{eqnarray}
where $\omega_1$ has been defined to equal the square root term.  Thus the radial motion is a superposition of two circular motions, the rapid modified cyclotron motion with frequency $\omega_{\mathrm{c}}^{\prime} = \omega_{\mathrm{c}}/2 + \omega_1$ and the usually much slower magnetron motion with a frequency $\omega_{\mathrm{m}} = \omega_{\mathrm{c}}/2 - \omega_1$.


\section{Laser cooling in the Penning trap}
\label{sec:lasercooling}

\begin{figure}
\begin{center}
\psfrag{y0}{$y_0$}
\psfrag{y}{$y$}
\psfrag{x}{$x$}
\includegraphics[width=0.8\textwidth]{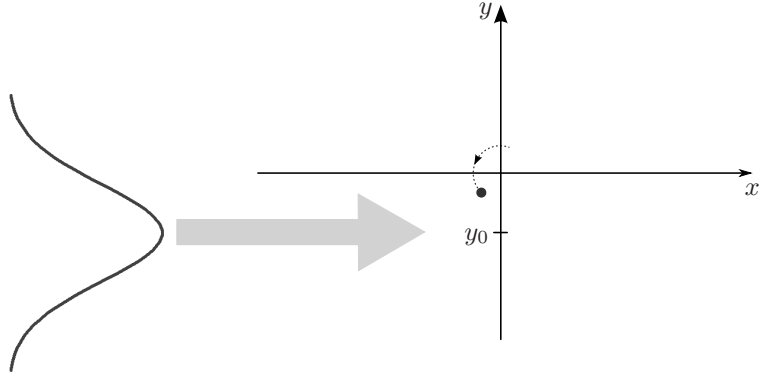}
\caption{Diagram showing the experimental situation assumed in the laser cooling model.  The Gaussian on the left shows the intensity of the beam as a function of $y$ and indicates the changing intensity experienced by the ion as it moves counter-clockwise in the trap.
\label{fig:coolingmodel}}
\end{center}
\end{figure}

The model presented in this paper is an extension of work first presented in~\cite{Thompson00}.  Only a brief description of the key elements is given here.  The model deals with the case of a single ion being Doppler cooled by a laser with a frequency $\nu_{\mathrm{L}}$, detuned from some strong transition at a frequency $\nu_0$.  The laser beam is directed parallel to the $x$ axis, but can be offset by an amount $y_0$ in the $y$ direction (figure~\ref{fig:coolingmodel}).  Absorption of photons from the laser beam leads to a net force on the ion in the $x$ direction.  Photon re-emission is assumed to occur in random directions and hence to cancel to zero.  Since the re-emission process is important in determining the ion temperatures achievable through laser cooling, it follows that the model as it stands cannot be used to calculate limiting temperatures.  Instead it yields the rates at which ion motions are laser cooled.

\begin{figure}
\begin{center}
\psfrag{y0}{$y_0$}
\psfrag{i2w}{$\sqrt{2}w$}
\psfrag{y}{$y$}
\psfrag{x}{$x$}
\psfrag{hdv}{$\lambda \delta_{\nu}$}
\psfrag{I}{$R$}
\psfrag{h(vlaser-v0)}{$\lambda \left(\nu_{\mathrm{l}} - \nu_0 \right)$}
\includegraphics[width=1.0\textwidth]{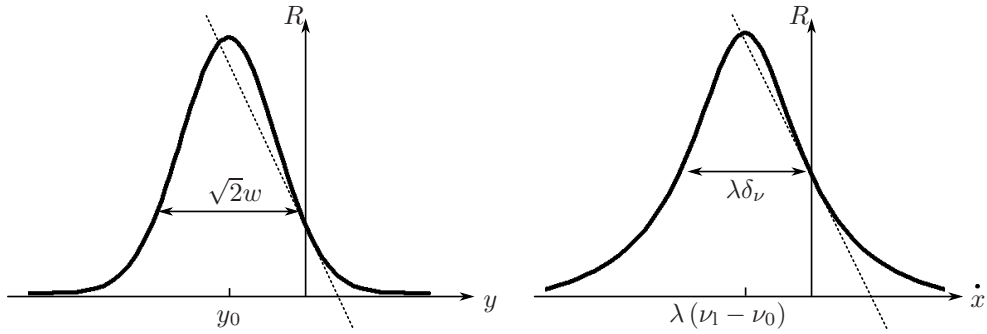}
\caption{\textbf{\textit{Left:}} If the amplitude of an ion's motion is small compared to the waist of the beam, the variation in scattering rate $R$ as a function of $y$ is approximately linear.  \textbf{\textit{Right:}} Similarly, if the ion velocity is small then the Doppler shift as the ion oscillates is small compared to the linewidth of the transition and so the change in scattering rate as a function of $\dot{x}$ is approximately linear.
\label{fig:coolingmodellinearapprox}}
\end{center}
\end{figure}

The Gaussian intensity profile of the beam means that there is a variation in the photon scattering rate, and therefore the force on the ion due to the laser, dependent on the position of the ion in the $y$ direction.  Similarly, the velocity of the ion in the $x$ direction leads to a Doppler shift of the laser frequency and hence the ion experiences different regions of the Lorentzian lineshape of the cooling transition as it moves in the trap.  The force on the ion is therefore dependent also on $\dot{x}$.  Providing the ion is cold, so that the amplitude of its motion and its velocity are small, the variation in photon scattering rate with both $y$ and $\dot{x}$ is approximately linear (figure~\ref{fig:coolingmodellinearapprox}).

The total force on the ion due to its interaction with the laser beam is in the $x$ direction and can be written as $F = F_0 + F_{\alpha} + F_{\beta}$.  Here $F_0$ is a constant that leads only to a slight offset in the equilibrium position of the ion and is henceforth ignored.  $F_{\alpha}$ and $F_{\beta}$ are the components of the force that are considered to vary linearly with ion position $y$ and velocity $\dot{x}$ respectively.  The constants of proportionality $\alpha$ and $\beta$ are defined such that $F_{\alpha} = -2 \alpha m y$ and $F_{\beta} = -2 \beta m \dot{x}$.  The values of $\alpha$ and $\beta$ are determined by calculating the slopes of the power-broadened scattering rate curves shown in figure~\ref{fig:coolingmodellinearapprox} at the points where $y$ and $\dot{x}$ are zero.  Note that $\alpha$ and $\beta$ can be changed by varying the beam offset, $y_0$, and frequency, $\nu_{\mathrm{L}}$, but the two parameters are not independent.

Adding these forces to the equations of motion obtained in section~\ref{sec:penningmotion} yields the equations of motion in the presence of laser cooling
\begin{eqnarray}
\label{eq:PenningcoolingxyDE}
\ddot{x} + \omega_{\mathrm{c}} \dot{y} + 2 \beta \dot{x} - \frac{\omega_{\mathrm{z}}^2}{2} x + 2 \alpha y = 0, \nonumber\\
\ddot{y} - \omega_{\mathrm{c}} \dot{x} - \frac{\omega_{\mathrm{z}}^2}{2} y = 0.
\end{eqnarray}
Assuming that $\beta \ll \omega_{\mathrm{c}}$ and $\alpha \ll \omega_{\mathrm{z}}^2 / 2$, corresponding physically to the assumption that the laser interaction is sufficiently weak that it does not significantly affect the ion oscillation frequencies, and changing as before to a complex variable $u = x + iy$, leads to
\begin{equation}
\label{eq:PenningcoolinguDE}
\ddot{u} + \left( \beta - i \omega_{\mathrm{c}} \right)  \dot{u} - \left( \frac{\omega_{\mathrm{z}}^2}{2} + i \alpha \right)  u = 0.
\end{equation}
Trying once again solutions corresponding to circular motion of the form $u=u_0\mathrm{e}^{i \omega t}$, where $\omega$ is now also complex, gives
\begin{equation}
\label{eq:PenningcoolinguDEsolution}
u_0 \mathrm{e}^{i \omega t} \bigg[ \omega^2 - \left( \omega_{\mathrm{c}} + i \beta \right)  \omega + \frac{\omega_{\mathrm{z}}^2}{2} + i \alpha \bigg]  = 0. \nonumber
\end{equation}
This has the roots
\begin{equation}
\label{eq:PenningcoolinguDEroots}
\omega = \frac{\omega_{\mathrm{c}} + i \beta}{2} \pm \sqrt{\omega_1^2 - \frac{\beta^2}{4} + \frac{i \omega_{\mathrm{c}} \beta}{2}  - i \alpha}. 
\end{equation}
A Taylor expansion of the square root term with the assumption that $\alpha \ll \omega_1^2$ and $\beta \ll \omega_1$ leads to the solutions
\begin{equation}
\label{eq:Penningcoolingmodcyc}
\omega = \omega_{\mathrm{c}}^{\prime} + i \frac{\beta \omega_{\mathrm{c}}^{\prime} - \alpha}{2 \omega_1}
\end{equation}
and
\begin{equation}
\label{eq:Penningcoolingmag}
\omega = \omega_{\mathrm{m}} + i \frac{\alpha - \beta \omega_{\mathrm{m}}}{2 \omega_1}.
\end{equation}
The frequencies of motion, given by the real part of $\omega$, are therefore the same as those obtained in the absence of laser cooling.  

Looking at the damping rates, given by the imaginary parts of the solutions, the difficulty in achieving efficient laser cooling in the Penning trap is immediately apparent.  Positive damping of the modified cyclotron motion requires that $\beta \omega_{\mathrm{c}}^{\prime} > \alpha$, whereas cooling of the magnetron motion requires that $\alpha > \beta \omega_{\mathrm{m}}$.  It should be remembered that $\alpha$ and $\beta$ are related by the overall scattering rate --- offsetting the beam further may increase the magnitude of $\alpha$, but will always reduce the overall scattering rate and therefore $\beta$.  This means that to obtain efficient cooling of the modified cyclotron motion a red-detuned laser (positive $\beta$) is needed, with perhaps a slight offset of the beam in the $y$ direction (negative $\alpha$).  Strong cooling of the magnetron motion requires a blue-detuned laser (negative $\beta$) with a small offset in the $-y$ direction.  These requirements are incompatible and therefore it is impossible to achieve \textit{efficient\/} laser cooling of both motions simultaneously in a conventional Penning trap.  Limited damping of both motions can, however, be achieved if the criterion $\omega_{\mathrm{m}} < \alpha / \beta < \omega_{\mathrm{c}}^{\prime}$ is satisfied, where both $\alpha$ and $\beta$ must be positive~\cite{Wineland82,Horvath99}.  This corresponds to a red-detuned laser and an offset of the beam such that it propagates in the same direction as the ions travel around the centre of the trap.  This is confirmed by numerical calculations of the cooling rates for the modified cyclotron and magnetron motions as a function of the beam offset $y_0$ and detuning $\nu_{\mathrm{L}} - \nu_0$, and by numerous experiments~\cite{Thompson86,Thompson88}. 

\begin{figure}
\begin{center}
\includegraphics[width=1.0\textwidth]{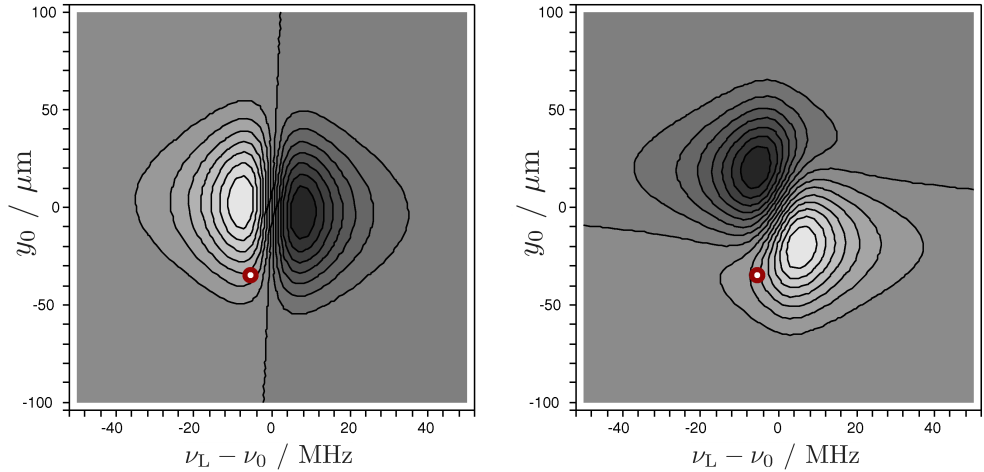}
\caption{Modified cyclotron \textit{(left)} and magnetron \textit{(right)} motion cooling rates calculated using the described cooling model.  A beam waist of 50 $\mathrm{\mu}$m and a peak intensity of $0.1 \times I_{\mathrm{sat}}$ is assumed.  Lighter shades indicate larger cooling rates and in each case the zero cooling rate contour is that which bisects the plot.  The small ring indicates a position at which some cooling of both motions can be achieved.
\label{fig:coolingmodelresults}}
\end{center}
\end{figure}

Figure~\ref{fig:coolingmodelresults} shows the results of such a calculation for a calcium ion with trapping parameters that are typical of those in recent experiments~\cite{Koo04}.  A beam waist of 50$\mu$m is assumed and the peak beam intensity used is 0.1 times the saturation intensity of the transition.  The only way to achieve cooling of both motions is to red-detune and offset the laser beam, as indicated in figure~\ref{fig:coolingmodelresults}.

\section{A model for axialization}
\label{sec:axialization}

The difficulty in efficiently laser cooling the magnetron motion can be eased by applying a weak radially quadrupolar electric potential that oscillates at the true cyclotron frequency, $\omega_{\mathrm{c}}$.  This is the sum of the two radial motional frequencies and hence the potential acts to couple these two components of the motion together.  The relative amplitudes of the two components are then linked so that the efficient cooling of the modified cyclotron motion also leads to a reduction in amplitude of the magnetron motion.  Note that the sum frequency, rather than the difference, is used because the magnetron motion is unstable --- energy must be added to this motion in order to reduce its amplitude.  This is due to the fact that the saddle point of $\phi$ in~(\ref{eq:Lorentz force}) has a maximum in the radial plane at the origin.  Throughout this paper we adopt a pragmatic definition of `cooling' of the magnetron motion as being a process that reduces its amplitude.

Axialization, or `radial centering', has been used in combination with resistive cooling~\cite{Werth03} and buffer gas cooling using both neutral and charged species~\cite{Savard91,Kellerbauer06}.  More recently it has been used with laser cooling~\cite{Powell02,Powell03}.  Outside of the laser cooling community the axialization technique is also known as `sideband cooling'.  This is not related to the laser cooling technique of the same name~\cite{Diedrich89}.  Theoretical descriptions of axialization have been given in~\cite{Wineland75,Wineland76,Brown86,Schweikhard93,Bollen95}.  The analytical model for laser cooling in the Penning trap presented in the previous section can be extended to include the effect of an axializing field, and doing so helps explain various effects observed in recent experiments. 

Consider a radially quadrupolar potential oscillating with a frequency $\omega_{\mathrm{a}}$,
\begin{equation}
\label{eq:axfieldlab}
\phi_{\mathrm{a}} = V_0 \frac{x^2 - y^2}{2 r_0^2} \cos \left( \omega_{\mathrm{a}} t \right).
\end{equation}
In polar coordinates this becomes
\begin{eqnarray}
\label{eq:axfieldpolar}
\phi_{\mathrm{a}} & = & \frac{V_0}{2} \frac{r^2}{r_0^2} \left( \cos^2 \theta - \sin^2 \theta \right) \cos \left( \omega_{\mathrm{a}} t \right) \nonumber \\
& = & \frac{V_0}{4} \frac{r^2}{r_0^2} \left( \cos \left( 2 \theta + \omega_{\mathrm{a}} t \right) + \cos \left( 2 \theta - \omega_{\mathrm{a}} t \right) \right),
\end{eqnarray}
which is equivalent to two counter-rotating quadrupolar potentials.  Transforming into a frame rotating at a frequency $\omega_{\mathrm{r}}$ by making the substitutions $\theta^{\prime} = \theta - \omega_{\mathrm{r}} t$ and $r^{\prime} = r$, means that in this frame
\begin{equation}
\label{eq:axfieldrot}
\phi^{\prime}_{\mathrm{a}} = \frac{V_0}{4} \frac{r^{\prime 2}}{r_0^2} \left( \cos \left( 2 \theta^{\prime} + \left( 2 \omega_{\mathrm{r}} + \omega_{\mathrm{a}} \right) t \right) + \cos \left( 2 \theta^{\prime} + \left( 2 \omega_{\mathrm{r}} - \omega_{\mathrm{a}} \right) t \right) \right).
\end{equation}
Clearly if $\omega_{\mathrm{r}} = \omega_{\mathrm{a}} / 2$ then one of these components is static and the other rotates with frequency $- 4 \omega_{\mathrm{r}}$.  It is anticipated that the oscillating quadrupolar potential will give rise to a coupling between motions only when close to resonance with a sum or difference of motional frequencies.  In a frame rotating at $\omega_{\mathrm{c}} / 2$ the two motional frequencies are $\pm \omega_1$, so the sum frequency is zero.  Thus the static component of $\phi_{\mathrm{a}}$ will be resonant, but the (rapidly) rotating component will in general not be.  It is therefore possible to ignore the rotating component and calculate the acceleration in the rotating frame due to the static component alone.  If $x^{\prime}$ and $y^{\prime}$ are cartesian coordinates in the frame rotating with a frequency $\omega_{\mathrm{r}} = \omega_{\mathrm{a}} / 2$ then
\begin{eqnarray}
\label{eq:axaccelerationpot}
\phi^{\prime}_{\mathrm{a}} & = & \frac {V_0}{4 r_0^2} r^{\prime 2} \cos\left( 2 \theta^{\prime} \right) \nonumber \\
& = & \frac {V_0}{4 r_0^2} \left( x^{\prime 2} - y^{\prime 2} \right).
\end{eqnarray}
The additional forces on a positively charged ion can then be calculated as
\begin{eqnarray}
\label{eq:axaccelerationforce}
F_{\mathrm{x}^{\prime}} & = & - \frac{e V_0}{2 r_0^2} x^{\prime} \nonumber \\
F_{\mathrm{y}^{\prime}} & = & + \frac{e V_0}{2 r_0^2} y^{\prime}.
\end{eqnarray}
Making the substitution $v = x^{\prime} + i y^{\prime}$ allows this to be written as
\begin{equation}
\label{eq:axaccelerationforcev}
F_{\mathrm{v}} = F_{\mathrm{x}^{\prime}} + i F_{\mathrm{y}^{\prime}} = - \frac{e V_0}{2 r_0^2} v^{\ast}.
\end{equation}
By introducing $\epsilon = e V_0 / 2 m r_0^2$, the force due to axialization is simply given by $- \epsilon m v^{\ast}$.  This force can now be included in the laser cooling model.  

Equation~(\ref{eq:PenningcoolinguDE}), which describes the acceleration due to laser cooling in the $u$ (laboratory) frame, is first transformed into the $v$ (rotating) frame in which the axializing force has been determined.  Using the relation $u = v \mathrm{e}^{i \omega_{\mathrm{r}} t}$ leads to
{\setlength\arraycolsep{0pt}
\begin{eqnarray}
\label{eq:PenningcoolingvDE}
\mathrm{e}^{i \omega_{\mathrm{r}} t} && \bigg[ \ddot{v} + \left( \beta - i \omega_{\mathrm{c}} + 2 i \omega_{\mathrm{r}} \right) \dot{v} \nonumber \\
&& \quad - \left( \frac{\omega_{\mathrm{z}}^2}{2} + i \alpha + \omega_{\mathrm{r}}^2 - \omega_{\mathrm{r}} \left( \omega_{\mathrm{c}} + i \beta \right) \right) v \bigg] = 0. 
\end{eqnarray}}
Adding in the axialization term then gives
{\setlength\arraycolsep{0pt}
\begin{eqnarray}
\label{eq:PenningaxializationvDE}
&&\ddot{v} + \left( \beta - i \omega_{\mathrm{c}} + 2 i \omega_{\mathrm{r}} \right) \dot{v} \nonumber \\
&& \phantom{\mathrm{e}^{i \omega_{\mathrm{r}} t}} \quad - \left( \frac{\omega_{\mathrm{z}}^2}{2} + i \alpha + \omega_{\mathrm{r}}^2 - \omega_{\mathrm{r}} \left( \omega_{\mathrm{c}} + i \beta \right) \right) v + \epsilon v^{\ast}  = 0.
\end{eqnarray}}
The appropriate trial solution now corresponds to elliptical motion with uniform decay, because the rotational symmetry is broken by the $v^{\ast}$ term.  This elliptical motion is a superposition of two circular motions in opposite senses with different amplitudes.  In general,
\begin{equation}
\label{eq:PenningaxializationvDEtrial}
v = A \mathrm{e}^{i \omega t} + B \mathrm{e}^{-i \omega^{\ast} t}.			\end{equation}
For each solution $\omega$, $A$ and $B$ are the counterclockwise and clockwise components of a single motion.  Note the use of $\omega^{\ast}$ in the second term --- the damping rate (real part of the exponent) must be the same for both components even though they rotate in different senses.

A complete solution to~(\ref{eq:PenningaxializationvDE}) using this trial solution can be found in Appendix~A.  In it we define $\Delta$ to be half the detuning of the axialization drive from the cyclotron frequency, so that $\omega_{\mathrm{a}} = \omega_{\mathrm{c}} + 2 \Delta$.  Thus the rotation frequency of the frame in which the axializing quadrupolar drive is static is $\omega_{\mathrm{r}} = \omega_{\mathrm{c}} / 2 + \Delta$.  We express the solutions $\omega$ in terms of the unperturbed motional frequencies $\omega_1$ and additional (real) frequency shift and damping terms $\delta_0$ and $\gamma_0$ such that $\omega = \omega_1 + \delta_0 + i \gamma_0$.  The solutions obtained are stated here as
\begin{equation}
\label{eq:Penningaxializationdeltaroots}
\delta_0 = \pm \frac{1}{\sqrt{2}} \sqrt{N + \sqrt{N^2 + \Delta^2 M^2}}
\end{equation}
and
\begin{equation}
\label{eq:PenningaxializationgammaM}
\gamma_0 = \frac{\beta}{2} + \frac{\Delta M}{2 \delta_0},
\end{equation}
where
\begin{equation}
\label{eq:M}
M = \left(2 \alpha - \beta \omega_{\mathrm{c}} \right) / 2 \omega_1
\end{equation}
and
\begin{equation}
\label{eq:N}
N = \Delta^2 - \frac{M^2}{4} + \frac{\left| \epsilon \right|^2}{4 \omega_1^2}.
\end{equation}
Note that the parameter $M$ is a measure of the strength of the laser cooling.  $N$, on the other hand, is dependent not only on the laser parameters, but also the amplitude and detuning of the axialization drive. 

Recalling that the motional frequencies in the laboratory frame are given by $\omega = \omega_{\mathrm{r}} \pm \left( \omega_1 + \delta_0 \right)$ and that $\omega_{\mathrm{r}} = \omega_{\mathrm{c}} / 2 + \Delta$ it is apparent that having two solutions for $\delta_0$ leads to components in the motion at four frequencies, given by
\begin{equation}
\label{eq:Penningaxializationfrequencies1}
\omega = \omega^{\prime}_{\mathrm{c}} + \Delta + \delta_0
\end{equation}
and
\begin{equation}
\label{eq:Penningaxializationfrequencies2}
\omega = \omega_{\mathrm{m}} + \Delta - \delta_0
\end{equation}
where $\delta_0$ takes the two values given by~(\ref{eq:Penningaxializationdeltaroots}).


\section{Results of the model}
\label{sec:resultsofmodel}

The behaviour predicted by the model depends on the relative values of the terms that make up the parameter $N$, defined by~(\ref{eq:N}).  Clearly if the amplitude of the axializing quad\-ru\-polar drive is set to zero then the model should yield unshifted frequencies in the laboratory frame, i.e.~$\omega_{\mathrm{c}}^{\prime}$ and $\omega_{\mathrm{m}}$.   This is confirmed by setting $\epsilon$ to zero.  In this case $N = \Delta^2 - M^2 / 4$ and~(\ref{eq:Penningaxializationdeltaroots}) reduces to
\begin{equation}
\label{eq:axializationfreqE=0}
\delta_0 = \pm \Delta. 
\end{equation}
In the laboratory frame the frequency shifts for each motion (given by $\delta_0 + \Delta$) are therefore $+ 2 \Delta$ or zero.  As we show in section~\ref{subsec:phase_behaviour}, the amplitudes of the former solution turn out to be zero, leaving us, as expected, with unshifted frequencies.  We can obtain the damping rates by putting $\delta_0 = \pm \Delta$ into~(\ref{eq:PenningaxializationgammaM}), giving
\begin{eqnarray}
\label{eq:axializationdampingE=0}
\gamma_0 & = & \frac{\beta \pm M}{2} \nonumber \\
& = & \frac{\beta \omega_1 \pm \left( \alpha - \frac{\beta \omega_{\mathrm{c}}}{2} \right)}{2 \omega_1}, 
\end{eqnarray}
so either
\begin{equation}
\label{eq:axializationdampingE=0a}
\gamma_0 = \frac{\alpha - \beta \omega_{\mathrm{m}}}{2 \omega_1} 
\end{equation}  
or
\begin{equation}
\label{eq:axializationdampingE=0b}
\gamma_0 = \frac{\beta \omega_{\mathrm{c}}^{\prime} - \alpha}{2 \omega_1}. 
\end{equation}  
For the modified cyclotron motion it is the latter solution that corresponds to the unshifted motional frequency in the laboratory frame.  The $-$ sign in~(\ref{eq:Penningaxializationfrequencies2}) means that the solutions for $\delta_0$ are reversed for the magnetron motion, and hence it is the former damping rate that applies to the unaltered frequency.  Thus the damping rate obtained for each motion is the same as that from the model without axialization.  

Interestingly, when there is some non-zero amplitude for the axial\-ising quadru\-polar drive, $\epsilon$, the behaviour of the system in terms of frequency shifts becomes dependent on the laser parameters.  The critical factor is the balance of the three terms in the expression for $N$.  Note that if the detuning of the axializing drive is made sufficiently large the $\Delta^2$ term in $N$ will always dominate and $\delta_0$ will reduce once again to $\pm \Delta$, as it did for the case of no drive at all.  This demonstrates that axialization is very much a resonant effect and the drive frequency must be close to the true cyclotron frequency if it is to have an effect.  The effective width of the resonance is related to the amplitude of the quadrupolar drive and the strength of the laser cooling.  Close to resonance $\Delta$ is small and the frequency behaviour depends on the relative magnitudes of $\epsilon / \omega_1$ and $M$.  We shall now consider in turn three cases corresponding to $\left(\epsilon / \omega_1 \right)^2 \ll M^2$, $\left(\epsilon / \omega_1 \right)^2 \sim M^2$ and $\left(\epsilon / \omega_1 \right)^2 \gg M^2$.

\subsection{Case 1: $\left(\epsilon / \omega_1 \right)^2 \ll M^2$}
\label{subsec:EllM}
\begin{figure}
\begin{center}
\begin{tabular}{cc}
\begin{minipage}{0.465\textwidth}
\begin{center}
(a)
\includegraphics[width=1.0\textwidth]{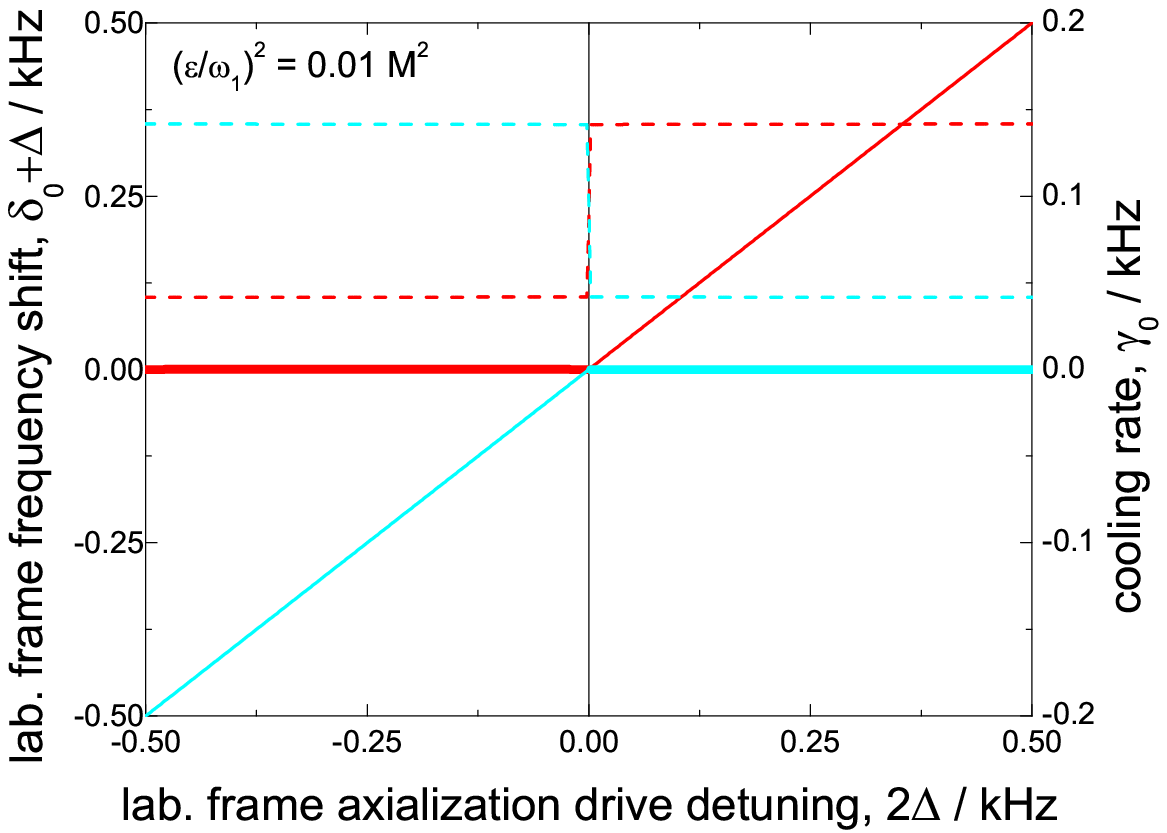}
\end{center}
\end{minipage}
& 
\begin{minipage}{0.465\textwidth}
\begin{center}
(b)
\includegraphics[width=1.0\textwidth]{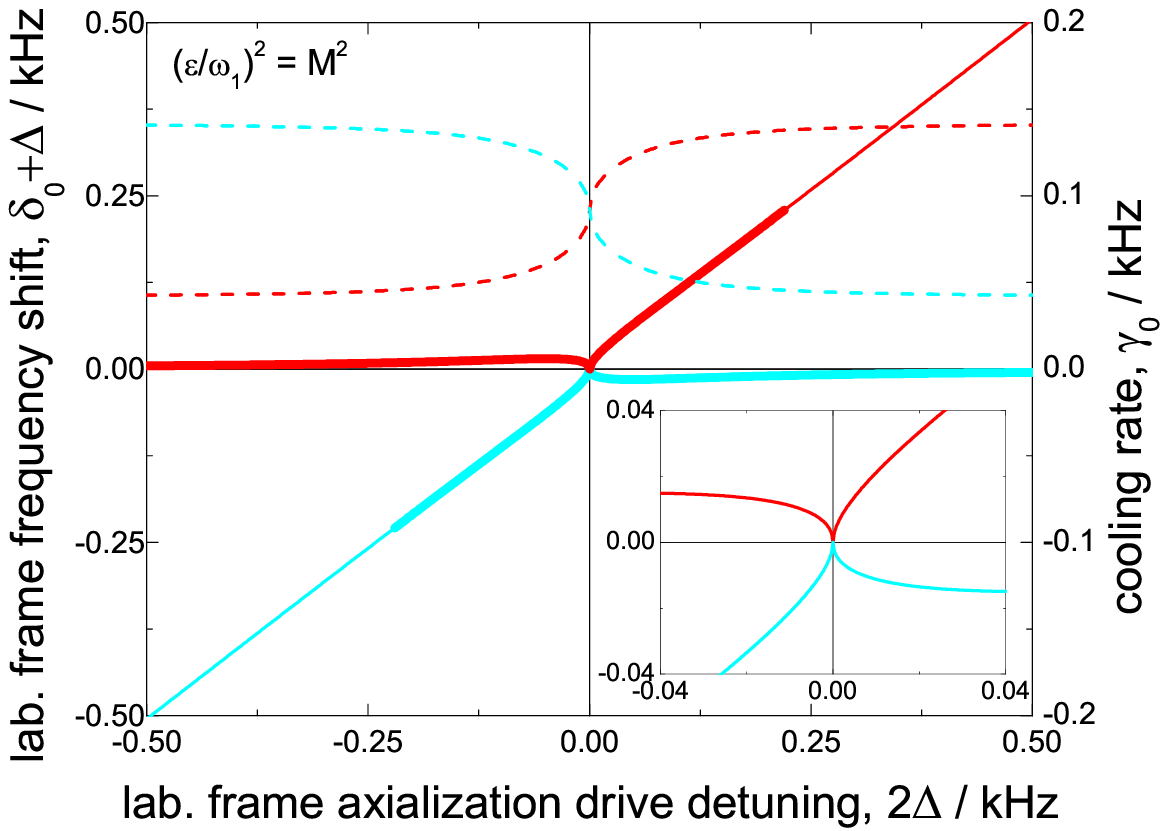}
\end{center}
\end{minipage}
\\
&
\\
&
\\
\begin{minipage}{0.465\textwidth}
\begin{center}
(c)
\includegraphics[width=1.0\textwidth]{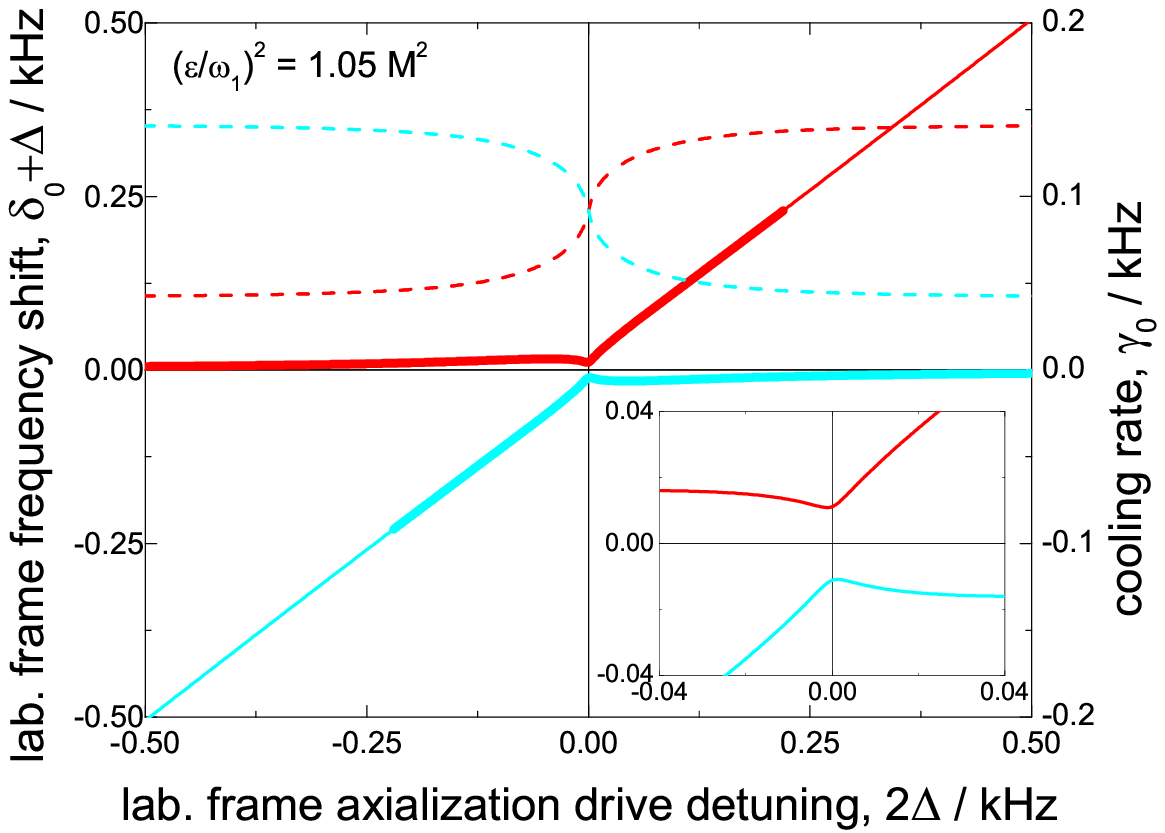}
\end{center}
\end{minipage}
& 
\begin{minipage}{0.465\textwidth}
\begin{center}
(d)
\includegraphics[width=1.0\textwidth]{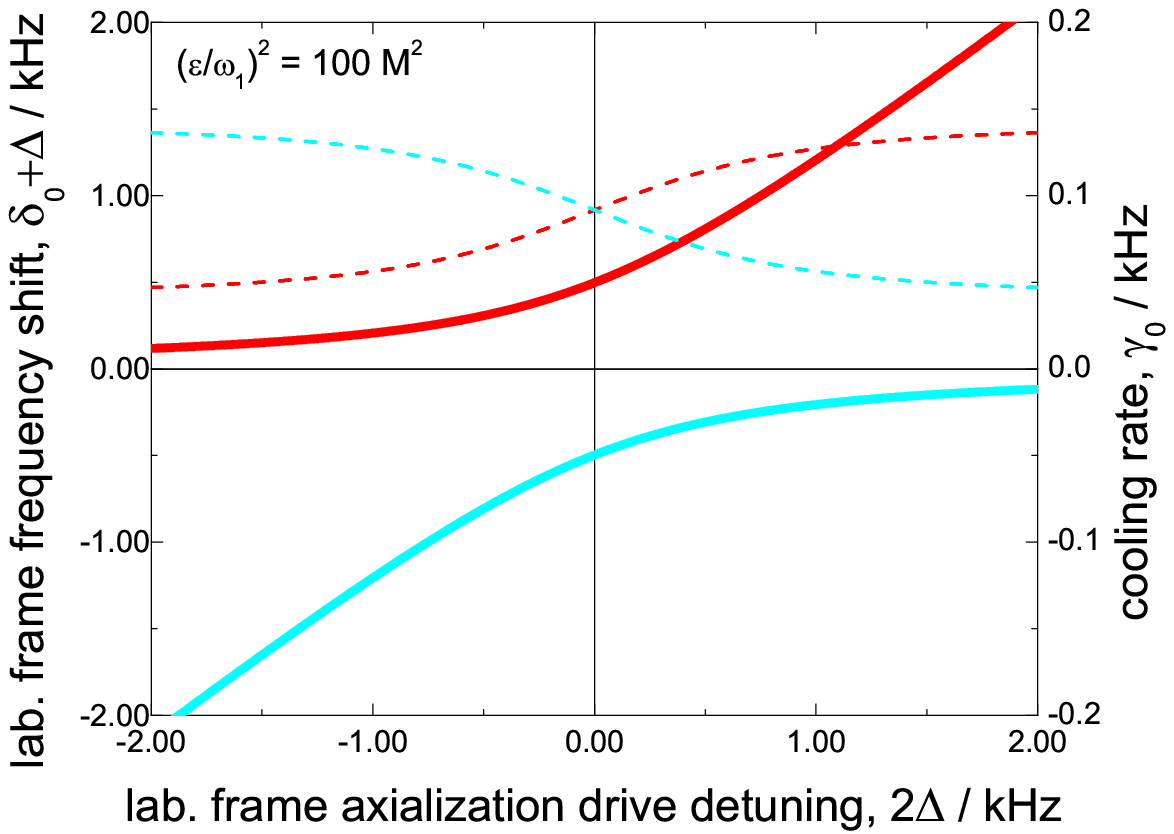}
\end{center}
\end{minipage}
\\
&
\end{tabular}
\caption{Motional frequency shifts (solid lines) and cooling rates (dashed lines) relative to the magnetron motion in the laboratory frame for the cases where \textbf{(a)} $\left(\epsilon / \omega_1 \right)^2 \ll M^2$, \textbf{(b)} $\left(\epsilon / \omega_1 \right)^2 = M^2$, \textbf{(c)} $\left(\epsilon / \omega_1 \right)^2 > M^2$ and \textbf{(d)} $\left(\epsilon / \omega_1 \right)^2 \gg M^2$.  The equivalent situation for the modified cyclotron motion differs only in that the cooling rate curves are swapped around.  Thin solid lines have been used to indicate when one of the components of the motion has less than half the amplitude of the other, as determined from an analysis described in section~\ref{subsec:phase_behaviour}.  Plotted parameters are for Ca$^+$ at 0.98 tesla, such that $\omega_c = 380 \times 2\pi$ kHz and $\omega_1 = 165 \times 2\pi$ kHz.  In each case $|M|^2=0.01$ kHz and $\alpha / \beta = 100 \times 2\pi$ kHz.  The inset plots are enlargements of the central region in which the behaviour for small axialization drive detuning can be more clearly seen.  Note the change in scale in plot (d).
\label{fig:axializationresults}}
\end{center}
\end{figure}

In this regime $N \sim \Delta^2 - M^2 / 4$ and the results are to good approximation the same as in the previously considered case of no axialization.  Calculated motional frequency shifts in the laboratory frame are plotted for the magnetron motion in figure~\ref{fig:axializationresults}(a), along with the corresponding damping (i.e.~laser cooling) rates.  The motion which has nearly unshifted frequency has the damping characteristics of the magnetron motion in an unperturbed Penning trap.  The other motion is the modified cyclotron motion `dressed' by the axializing field.  It has the large damping characteristic of the modified cyclotron motion but, as we shall see in section~\ref{subsec:phase_behaviour}, we expect its amplitude in this regime to be very small.  There is no visible coupling of the two damping rates.  Note the swapping of the curves at $\Delta = 0$.  The differently shaded curves correspond to taking either the positive or negative solution for $\delta_0$ and there is a sudden change in which root represents which motion.  Note also that a similar plot would be seen near the frequency of the modified cyclotron motion, in which an equivalent additional frequency would be seen corresponding to the `dressed' magnetron motion.

\subsection{Case 2: $\left(\epsilon / \omega_1 \right)^2 \sim M^2$}
\label{subsec:EsimM}
In this regime more interesting effects begin to occur.  Although the $\Delta^2$ term always dominates the expression for $N$ (except for very small $\Delta$), it is not always the dominant term in the equation for $\delta_0$.  If $N = \Delta^2$ then
\begin{equation} 
\label{eq:axializationfreqE=Ma}
\delta_0 = \pm \frac{1}{\sqrt{2}} \sqrt{\Delta^2 + \sqrt{\Delta^2 \left( \Delta^2 + M^2 \right)}}. 
\end{equation}  
In the region where $\Delta$ is small compared to $M$ this expression becomes
\begin{eqnarray}
\label{eq:axializationfreqE=Mb}
\delta_0 & = & \pm \frac{1}{\sqrt{2}} \sqrt{\Delta \left( \Delta + \sqrt{ \Delta^2 + M^2} \right) } \nonumber \\
& \approx & \pm \sqrt{\left|\frac{\Delta M}{2}\right|}.
\end{eqnarray}  
Therefore for small $\Delta$ the shift in motional frequencies will vary as $\pm \sqrt{\Delta}$ in the rotating frame.  As $\Delta$ becomes larger compared with $M$ the curves will again tend to $\pm \Delta$.  

Figure~\ref{fig:axializationresults}(b) is a plot of the frequency shifts and cooling rates calculated with $\left(\epsilon / \omega_1 \right)^2 = M^2$.  With the exception of $\epsilon$, all the parameters are the same as those used to plot figure~\ref{fig:axializationresults}(a).  The $\sqrt{\Delta}$ behaviour near resonance that was predicted to occur in the rotating frame is modified by the transformation into the laboratory frame.  Further from resonance the frequencies tend to those observed in the low axialization amplitude regime described in section~\ref{subsec:EllM}.  Note that whenever $\left(\epsilon / \omega_1 \right)^2 \le M^2$ the frequency shift $\delta_0$ is equal to zero when the axialization drive detuning is zero.

\begin{figure}
\begin{center}
\includegraphics[width=0.6\textwidth]{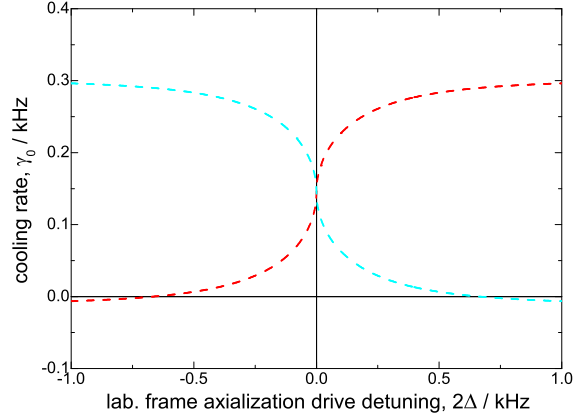}
\caption{Cooling rates for the magnetron motion when the ratio $\alpha / \beta$ is chosen to be $10 \times 2\pi$ kHz --- ten times smaller than in figure~\ref{fig:axializationresults}.  Although the mean cooling rate is larger near resonance, if the axialization drive is detuned by more than about 0.7 kHz one of the cooling rates becomes negative.  As in figure~\ref{fig:axializationresults}(b), $\left(\epsilon / \omega_1 \right)^2 = M^2$ with both terms now being 0.1 kHz$^2$.  As in the previous figures, $\omega_c = 380 \times 2\pi$ kHz and $\omega_1 = 165 \times 2\pi$ kHz.
\label{fig:largebetacoolingrates}}
\end{center}
\end{figure}

Looking at the damping rate curves, the useful property of the axialization technique is seen for the first time.  If $\left(\epsilon / \omega_1 \right)^2$ is nearly equal to or greater than $M^2$, then near resonance the damping rates of the two motions converge and there is an intermediate damping rate for both motions.  The benefit of coupling together the magnetron and modified cyclotron motions like this is not just that the magnetron damping is increased at the expense of the modified cyclotron damping, but that it is no longer necessary to maintain a large beam offset parameter $\alpha$ and small frequency gradient parameter $\beta$ in order to prevent heating of the magnetron motion.  The average cooling rate of the two motions is 
\begin{eqnarray} 
\label{eq:axializationavedampingE=M}
\gamma_{\mathrm{ave}} & = & \frac{1}{2} \left( \frac{\beta}{2} \pm \frac{\Delta M}{2 \delta_0} + \frac{\beta}{2} \mp \frac{\Delta M}{2 \delta_0} \right) \nonumber \\
& = & \frac{\beta}{2}.
\end{eqnarray}  
The $\alpha$ dependence has gone and the optimum case is represented by maximising $\beta$.  Furthermore, reducing the beam offset increases the beam intensity seen by the ions and so increases the overall scattering rate and hence $\beta$, yielding still higher mean cooling rates.  The only disadvantage to this is that it is possible to realise a situation in which one of the cooling rates is negative in the absence of axialization, and so the frequency of the axialization drive becomes even more important.  If the laser cooling is set to give large average damping rates but the axialization drive is not properly tuned to resonance, then the size of the magnetron orbit will be rapidly increased.  This is highlighted in figure~\ref{fig:largebetacoolingrates}, which shows the cooling rate curves for the situation in which $\beta$ is slightly greater than in figure~\ref{fig:axializationresults}.  The ratio $\alpha / \beta$ is now $10 \times 2\pi$ kHz so that the cooling criterion in the absence of axialization, $\omega_{\mathrm{m}} < \alpha / \beta < \omega_{\mathrm{c}}^{\prime}$, is no longer satisfied.  Clearly the damping rates when the axialization drive is near resonance are larger, but if the drive is detuned by more than about 0.7 kHz then the damping rate of one component of the motion becomes negative.

For the special case where $\left(\epsilon / \omega_1 \right)^2$ is exactly equal to $M^2$, the shapes of the cooling rate curves near resonance are given by substituting $\delta_0 = \pm \sqrt{ \left| \Delta M / 2 \right| }$ into the equation for $\gamma_0$.  This yields
\begin{equation} 
\label{eq:axializationdampingE=M}
\gamma_0 \approx \frac{\beta}{2} \pm \sqrt{\left|\frac{\Delta M}{2}\right|}. 
\end{equation}

\subsection{Case 3: $\left(\epsilon / \omega_1 \right)^2 \gg M^2$}
\label{subsec:EggM}
In this regime there is a similar damping rate coupling to that seen in the previous case, although the range of detunings in which the effect is significant is larger.  What is different, however, is that as soon as $\left(\epsilon / \omega_1\right)^2 > M^2$ the frequency shifts no longer cross at zero.  Instead there is an avoided crossing --- the frequency shifts approach each other as before for large $\Delta$, but near resonance they split apart.  The damping rates still swap over at $\Delta = 0$ as required, though at no point do the motions share the same frequency.  If $\left(\epsilon / \omega_1 \right)^2$ is only slightly greater than $M^2$ then there is an abrupt change in the gradients of the frequency shifts near the avoided crossing.  An example of this is given in figure~\ref{fig:axializationresults}(c), which shows the frequency shifts and cooling rates calculated using the same parameters as in the previous cases, but with $\left(\epsilon / \omega_1 \right)^2 = 1.05 M^2$.  As $\left(\epsilon / \omega_1 \right)^2$ is increased further the frequencies of the two components of the motion split further apart and the gradients of the frequency shifts vary more smoothly.  This case is illustrated in figure~\ref{fig:axializationresults}(d), in which the frequency shifts and cooling rates are shown for the case where $\left(\epsilon / \omega_1 \right)^2 = 100 M^2$.

If $\left(\epsilon / \omega_1 \right)^2 \gg M^2$ then the approximation $N = \Delta^2 + \epsilon^2 / \left( 4 \omega_1^2 \right)$ can be made.  The frequency shifts in the rotating frame are then given by
\begin{eqnarray} 
\label{eq:axializationfreqEggM}
\delta_0 & \approx & \pm \frac{1}{\sqrt{2}} \sqrt{N + \sqrt{N^2}} \nonumber \\
& = & \pm \sqrt{N} \nonumber \\
& = & \pm \sqrt{\Delta^2 + \frac{\epsilon^2}{4 \omega_1^2}}. 
\end{eqnarray}
The frequency shifts at $\Delta = 0$ are simply equal to $\epsilon / \left( 2 \omega_1 \right)$ and the hyperbolic form of this function is in excellent agreement with recent experimental measurements. 

The situation discussed here, where the laser cooling forces are weak compared to the forces due to the axialization drive, has been considered before using a different approach~\cite{Powell02,Powell03}.  The physical interpretation is that there is a continual exchange of amplitude back and forth between the two motions.  Thus if the amplitude of one of the motions can be reduced by laser cooling then with axialization the other motion will also be damped.

Avoided crossings resulting from the coupling together of motional modes of ions have been reported before by Cornell \textit{et~al.\/}~\cite{Cornell90} and H\"{a}ffner \textit{et~al.\/}~\cite{Werth03} using a Fourier Transform Ion Cyclotron Resonance technique.  Both groups coupled a radial motion to the axial motion in an attempt to extend resistive cooling of ions to the radial motions.  By a different method Cornell \textit{et~al.\/}~derived the same equation formulated here for the frequency shifts in the regime where the coupling rate is stronger than the cooling rate.  The result here extends theirs into the region of weaker coupling.  Furthermore the work by Cornell \textit{et~al.\/}~did not consider the phase behaviour of the ions' motion.  This motional phase behaviour forms an integral part of experimental techniques often used to measure ion cooling rates.  Since such measurements make it possible to determine the efficacy of axialization experimentally, we shall now investigate the motional phase behaviour in this model.  We shall also look at the relative amplitudes of the different motions by considering their ability to couple to an external driving field.


\subsection{Amplitude and phase response to a driving field}
\label{subsec:phase_behaviour} 

Thus far the motional frequencies and damping rates for an ion being laser cooled in the Penning trap in the presence of axialization have been calculated.  It was found, however, that the extra motional frequencies arising from the axializing field exist as solutions even if the amplitude of the drive, $\epsilon$, is zero and it was suggested that in this particular case these components of the motions would always have zero amplitude.  The solutions obtained (in the rotating frame) were of the form
\begin{equation}
\label{eq:PenningaxializationvDEtrialrepeat}
v = A \mathrm{e}^{i \omega t} + B \mathrm{e}^{-i \omega^{\ast} t}. 
\end{equation}
The real and imaginary parts of $\omega$ correspond to the motional frequency and damping rate respectively.  The relative amplitudes of the two components of each motion can be determined by finding an expression for the ratio of $A$ to $B$.  It is, however, interesting to derive a more general result that also includes the effect of an additional external force on the ion, since such a force is used in an experimental technique for measuring laser cooling rates~\cite{Eijkelenborg99:coolingrates}.

We consider an additional dipolar field oscillating at a frequency close to that of one of the characteristic ion motions in the laboratory frame, so that the ion is driven at this frequency.  The amplitude of the motion depends on how closely the driving frequency is tuned to one of the ion's natural motional frequencies.  When the drive frequency is exactly resonant with a natural frequency of the ion's motion the amplitude of the response can be extremely large.  The phase of the ion's motion relative to the driving field also depends on the drive frequency --- the ion will change smoothly from moving in phase with the drive to miving in antiphase with it as the drive is tuned from below to above resonance.   The width of this resonance in the response of an ion to a driving field is directly related to the damping present in the system.  Measurement of this width therefore allows the experimental determination of laser cooling rates~\cite{Eijkelenborg99:coolingrates}.

A complete solution to the equations of motion of a driven, laser-cooled ion in the presence of axialization is given in Appendix~B.  It adds to~(\ref{eq:PenningaxializationvDE}), the equation of motion for the ion, an additional driving force term $F \mathrm{e}^{i \omega t}$.  Using the trial solution given in~(\ref{eq:PenningaxializationvDEtrialrepeat}), where $\omega$ is now the frequency of the excitation drive in the rotating frame, the motional amplitudes $A$ and $B$ can be determined as
\begin{equation}
\label{eq:drivenPenningaxializationvDEsolution3a}
A = \frac{F C_{\mathrm{c}}}{C_{\mathrm{m}} C_{\mathrm{c}} - \left| \epsilon \right|^2},
\end{equation}
\begin{equation}
\label{eq:drivenPenningaxializationvDEsolution3b}
B = \frac{F \epsilon}{\left| \epsilon \right|^2 - C^{\ast}_{\mathrm{m}} C^{\ast}_{\mathrm{c}}},
\end{equation}
where
\begin{equation}
\label{eq:Cm}
C_{\mathrm{m}} = -2 \delta \omega_1 + i \beta \omega_1 -  2 \Delta \omega_1 - i \alpha + i  \beta \frac{\omega_{\mathrm{c}}}{2},
\end{equation}
\begin{equation}
\label{eq:Cc}
C_{\mathrm{c}} = -2 \delta \omega_1 + i \beta \omega_1  +  2 \Delta \omega_1 + i \alpha - i \beta \frac{\omega_{\mathrm{c}}}{2}.
\end{equation}
Both $A$ and $B$ are therefore functions of the amplitude and detuning of the excitation drive ($F$ and $\delta$, in the rotating frame) and the axialization drive ($\epsilon$ and $2 \Delta$), the laser parameters ($\alpha$ and $\beta$) and the trap frequencies ($\omega_{\mathrm{c}}$ and $\omega_1$).  Note that the ratio $A/B$ is independent of the driving force amplitude, $F$, and remains the same as $F$ tends to zero.  As expected we find that in the absence of axialization, when $\epsilon = 0$, the amplitude of one of the components of the motion, $B$, is also zero.

\begin{figure}
\begin{center}
\begin{tabular}{cr}
\begin{tabular}{cc}
\multicolumn{2}{c}{$\mathbf{(\epsilon / \omega_1)^2  = 0.01}\bi{M}^{\mathbf{2}}$}
\\
\begin{minipage}{0.37\textwidth}
\begin{center}
$\mathbf{|} \bi{A} \mathbf{|}$
\includegraphics[width=0.9\textwidth]{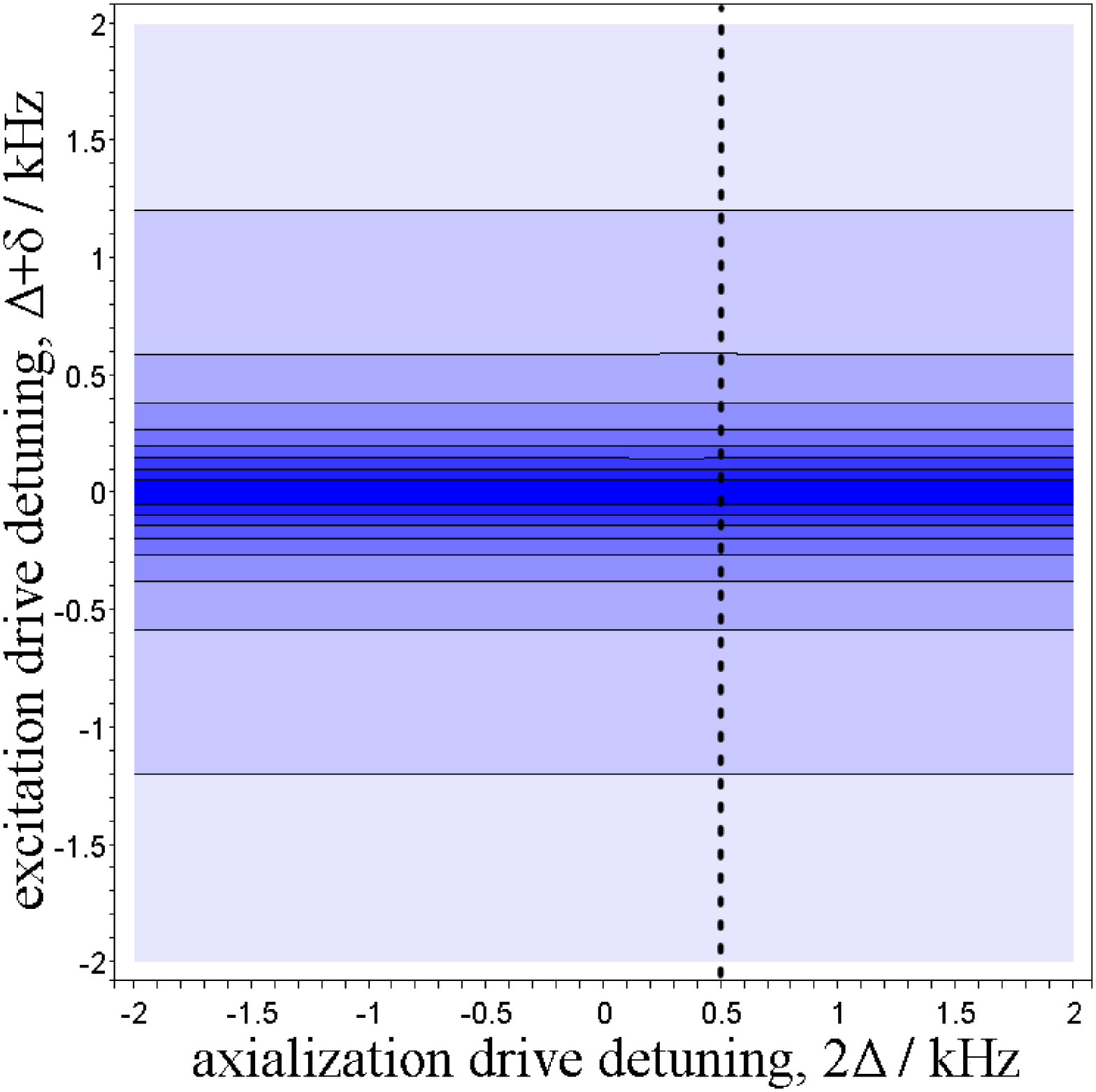}
\end{center}
\end{minipage}
& 
\begin{minipage}{0.37\textwidth}
\begin{center}
\textbf{Arg}$\mathbf{(}\bi{A}\mathbf{)}$
\includegraphics[width=0.9\textwidth]{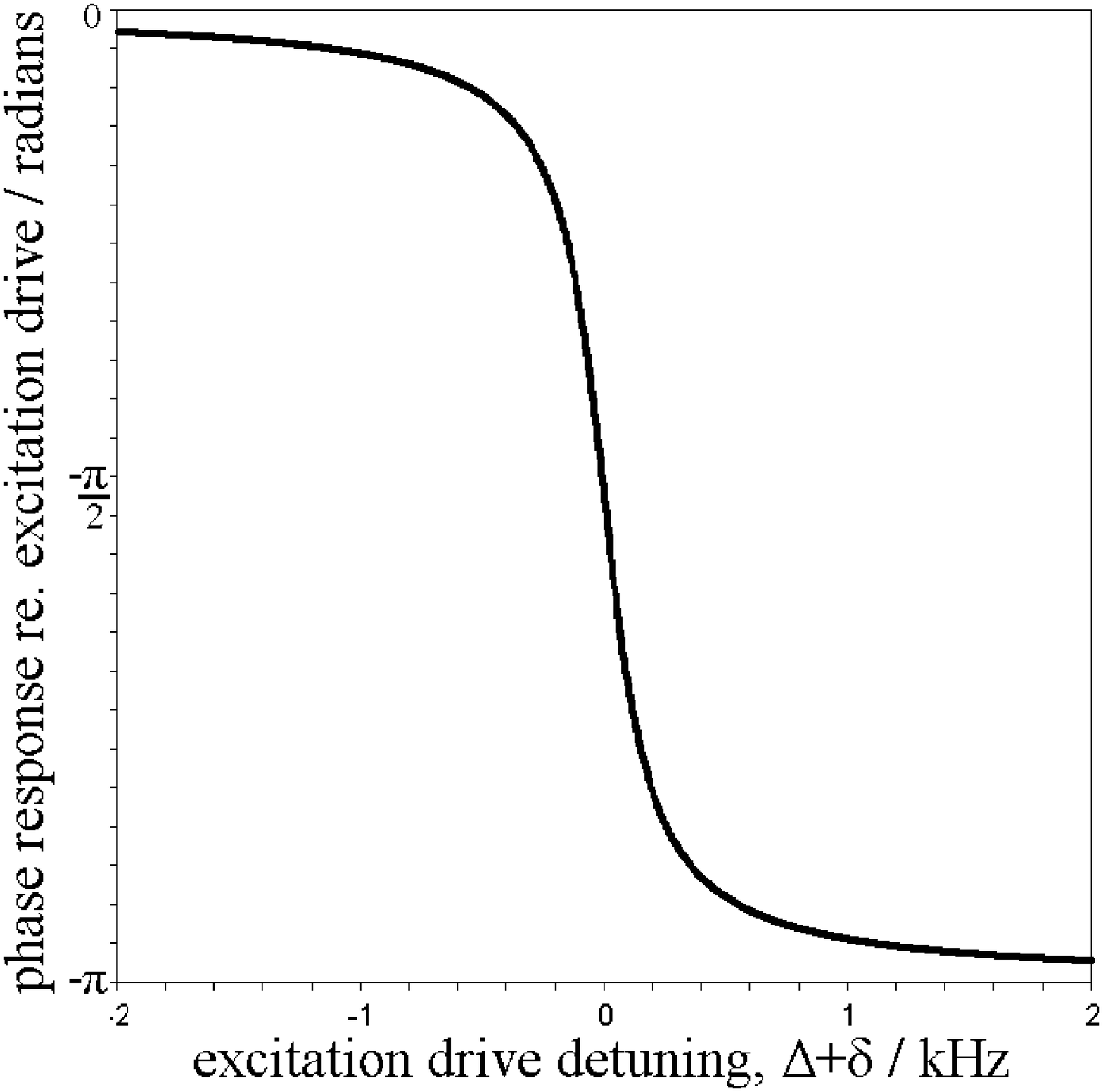}
\end{center}
\end{minipage}
\\
&
\\
\multicolumn{2}{c}{$\mathbf{(\epsilon / \omega_1)^2  = 100}\bi{M}^{\mathbf{2}}$}
\\
\begin{minipage}{0.37\textwidth}
\begin{center}
$\mathbf{|} \bi{A} \mathbf{|}$
\includegraphics[width=0.9\textwidth]{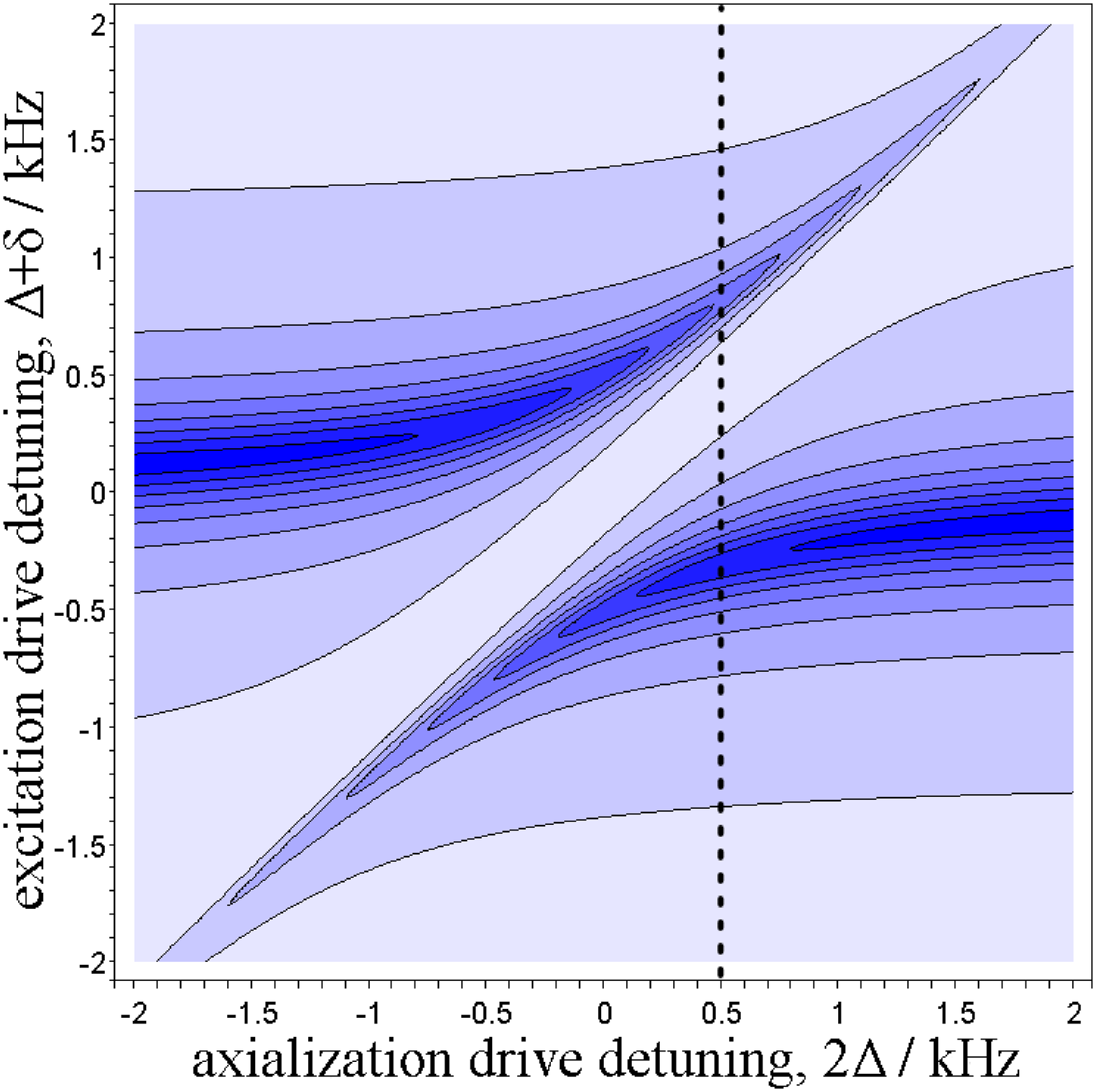}
\end{center}
\end{minipage}
& 
\begin{minipage}{0.37\textwidth}
\begin{center}
\textbf{Arg}$\mathbf{(}\bi{A}\mathbf{)}$
\includegraphics[width=0.9\textwidth]{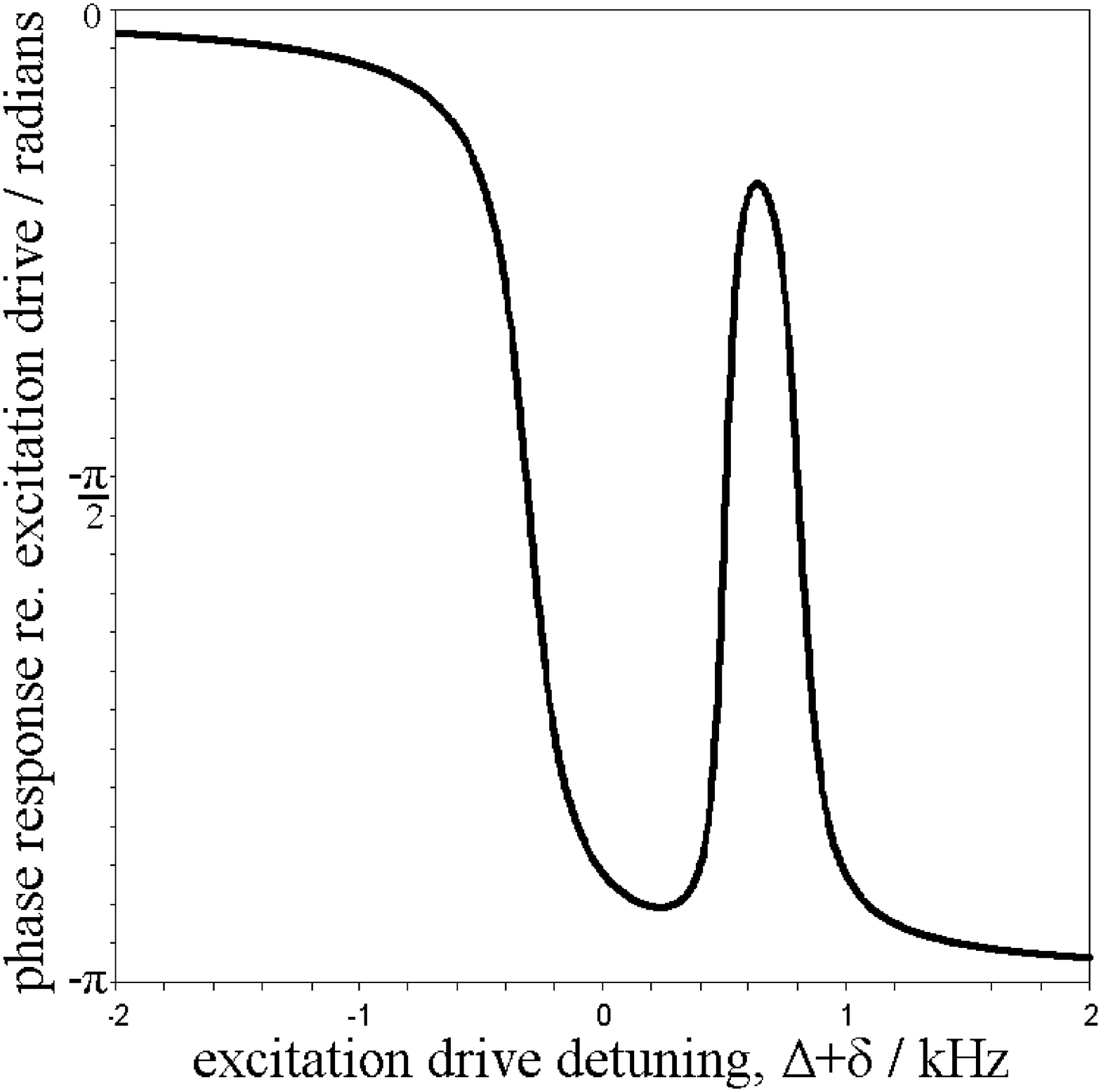}
\end{center}
\end{minipage}
\\
&
\\
\multicolumn{2}{c}{$\mathbf{(\epsilon / \omega_1)^2  = 100}\bi{M}^{\mathbf{2}}$}
\\
\begin{minipage}{0.37\textwidth}
\begin{center}
$\mathbf{|} \bi{B} \mathbf{|}$
\includegraphics[width=0.9\textwidth]{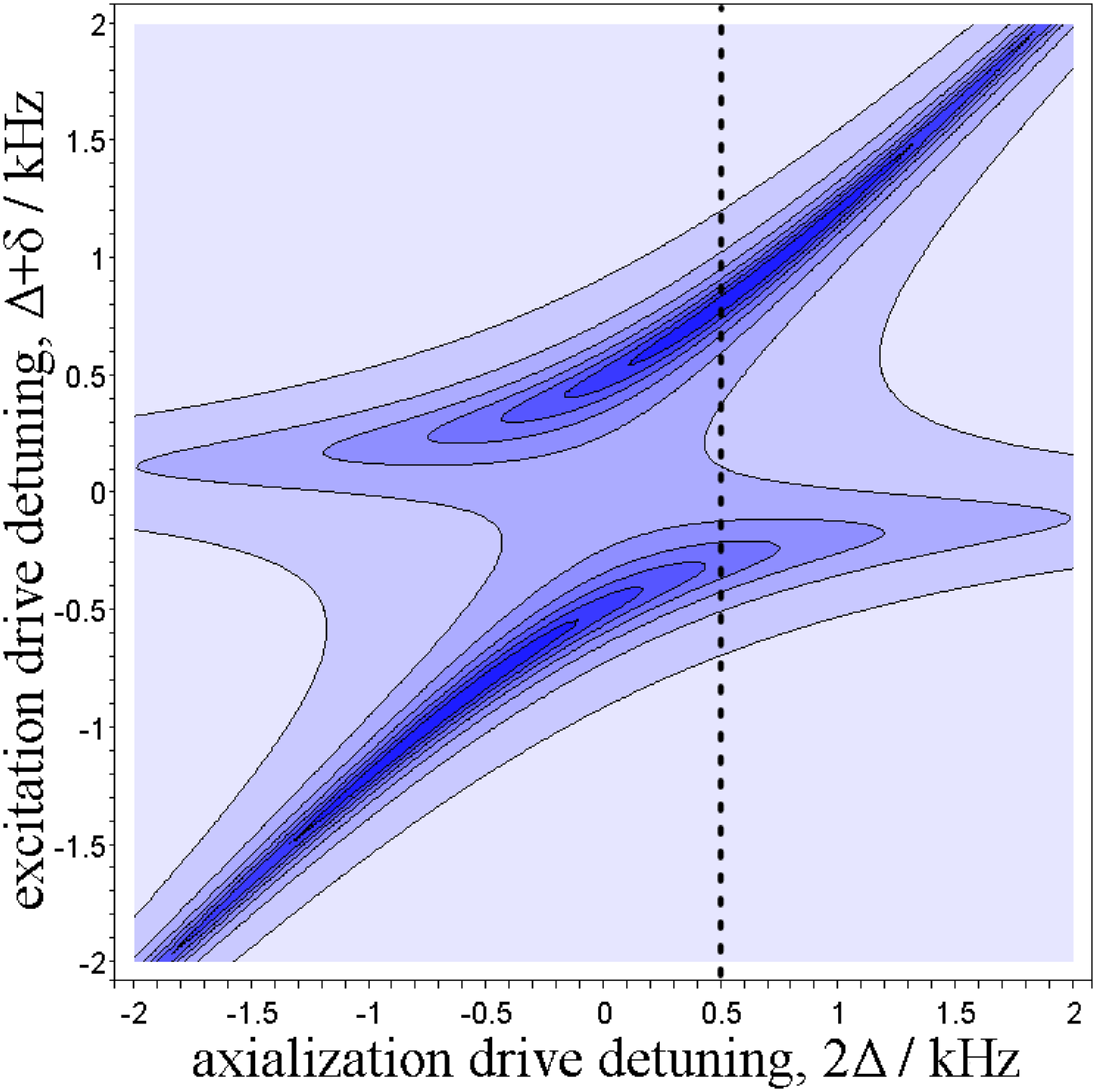}
\end{center}
\end{minipage}
& 
\begin{minipage}{0.37\textwidth}
\begin{center}
\textbf{Arg}$\mathbf{(} \bi{B} \mathbf{)}$
\includegraphics[width=0.9\textwidth]{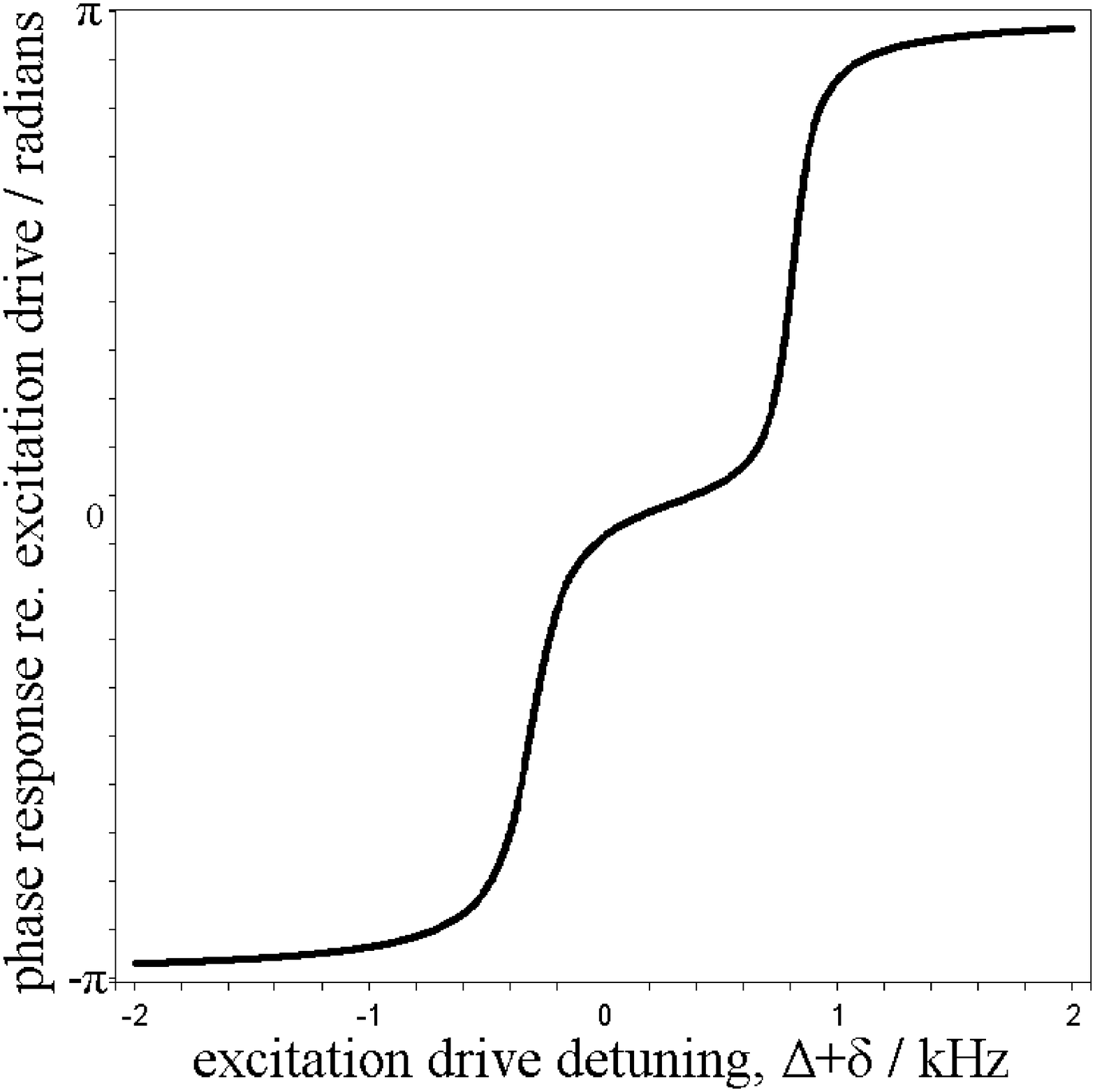}
\end{center}
\end{minipage}
\end{tabular}
&
\begin{minipage}{0.10\textwidth}
\begin{center}
\includegraphics[width=1.0\textwidth]{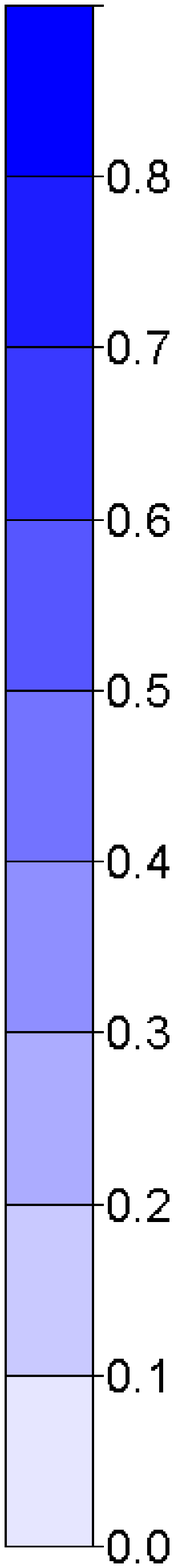}
\end{center}
\end{minipage}
\end{tabular}
\\
\caption{\textbf{\textit{Left:}} Contour plots showing the amplitude response of a calcium ion to an external dipolar drive near the magnetron frequency when $(\epsilon / \omega_1)^2  = 0.01 M^2$ and $(\epsilon / \omega_1)^2  = 100 M^2$, corresponding to cases (a) and (d) in figure~\ref{fig:axializationresults}.  $|A|$ is the response of the component of the motion being directly driven, whilst $|B|$ is the response of the indirectly driven component.  Note that the plot of $|B|$ is not shown for the case when $(\epsilon / \omega_1)^2  = 0.01 M^2$, since it is very close to zero for all values.  The scale is the same for each plot, but the absolute amplitudes are dependent on the driving force and are therefore arbitrary.  \textbf{\textit{Right:}} Plots of the phase of each component of the ion motion relative to the excitation drive.  The parameters used correspond to the dotted line on the contour plots.  In each case $\omega_c = 380 \times 2\pi$ kHz, $\omega_1 = 165 \times 2\pi$ kHz, $\alpha / \beta = 100 \times 2\pi$ kHz, and $|M|^2 = 0.01$~kHz.
\label{fig:axampphaseresponse}}
\end{center}
\end{figure}

$A$ and $B$ are complex quantities and the amplitude and phase response of the ion's motion relative to the excitation drive is obtained by taking their moduli and arguments respectively.  Figure~\ref{fig:axampphaseresponse} shows a typical contour plot of the amplitude response $|A|$ as a function of axialization and excitation drive detuning (in the laboratory frame) for a very small axialization amplitude.  The plotted parameters are identical to those used in figure~\ref{fig:axializationresults}(a).  This plot corresponds to the response of the motion being driven, and hence is the response that can be observed experimentally.  It is clearly evident from the plot that $|A|$ is only large when the excitation drive detuning is close to zero, and there is no indication of a second resonance frequency.  An equivalent plot of $|B|$ would correspond to the response of the component of the motion that is not being directly driven (i.e.\ the counter-rotating component) and would be difficult to measure experimentally.  When the axialization amplitude is very small, $|B|$ is found to be almost zero regardless of the axialization and excitation drive detunings.  Essentially this confirms that if there is no coupling between the two motions then driving one of them will have no effect upon the other.  These results validate the earlier assertion that the components of the motion with the extra frequencies observed in the model have very low amplitudes when the axialization drive amplitude is small or its detuning is large.

Also shown in figure~\ref{fig:axampphaseresponse} is Arg$(A)$, the motional phase of the driven component of the motion relative to the excitation drive.  The parameters used for this plot correspond to the dotted line marked on the contour plot of $|A|$.  There is a transition from being in phase with the excitation drive to lagging $\pi$ radians behind it as the drive is tuned through resonance.  This is identical to the phase response of a single classical oscillator relative to a driving force and is in agreement with measurements made in the absence of axialization~\cite{Eijkelenborg99:coolingrates}.

Figure~\ref{fig:axampphaseresponse} shows the amplitude response of the ion's motion when the axialization drive amplitude is large.  The parameters used to generate these plots are the same as those used in figure~\ref{fig:axializationresults}(d).  If the axialization drive is nearly resonant then the motion can be driven at two distinct frequencies.  The amplitude response $|A|$ to the excitation drive at each of these frequencies is similar --- as expected for a strongly coupled system.  Furthermore, driving one component of the motion, $A$, leads to a build up of amplitude in the other component, $B$, when close to resonance.  This is a direct indication of the efficacy of the axialization technique, since a damping force on one component of the motion will likewise lead to a reduction in amplitude of the other component.

The phases of the two components of the ion's motion relative to the excitation drive, given by Arg$(A)$ and Arg$(B)$, are also plotted in figure~\ref{fig:axampphaseresponse}.  Once again the parameters used are indicated by the dotted lines on the corresponding amplitude contour plots.  In addition to the $\pi$ phase shift observed in the case of weak axialization, Arg$(A)$ now exhibits a second phase shift at the position of the second resonance frequency.  The size of this second phase shift is found to tend towards $\pi$ as the axialization amplitude is increased.  

Arg$(B)$ also undergoes two phase shifts.  Unlike the component of the motion $A$, which always lags behind the excitation drive and so always has a relative phase between $-\pi$ and zero, the relative phase of $B$ can take any value.

The predicted phase behaviour of the measurable component of ion motion, Arg$(A)$, is in qualitative agreement with recent experimental results.


\section{Discussion}
\label{sec:discussion}

We have presented an analytical model for the axialization of laser-cooled ions in a Penning trap.  This model allows the calculation of the frequencies of motion of the ions and the behaviour is found to be different in each of three distinct regimes corresponding to weak, intermediate and strong coupling due to axialization.  The first and last of these regimes are easily accessed experimentally and the results of the model are in good agreement with measured frequencies.  The model can also be used to study the amplitude and phase response of the axialized ions relative to an oscillating dipolar electric potential, yielding results that are in qualitative agreement with measured data.

The model demonstrates that axialization can be used as a tool to enhance the efficiency of laser cooling of ions trapped in Penning traps and allows the calculation of the range of parameters for which such an improvement may be obtained.  By helping to overcome this significant experimental difficulty, axialization can make the Penning trap a viable alternative to the \textrm{RF} trap for quantum information processing and other experiments where precise control of a single trapped atomic particle is required.
\section*{Acknowledgements}
The authors gratefully acknowledge the financial support given by the EPSRC under grant GR/R14415/01 and the European Commission through its QUEST Research Training Network.
\appendix
\section{Solution of the equations for laser cooling in the presence of axialization}
\label{sec:appendixA}

Substituting trial solution~(\ref{eq:PenningaxializationvDEtrial}) into~(\ref{eq:PenningaxializationvDE}) yields
{\setlength\arraycolsep{0pt}
\begin{eqnarray}
\label{eq:PenningaxializationvDEsolution1}
\quad \quad \mathrm{e}^{i \omega t} && \bigg[ A \omega^2 - A ( i \beta + \omega_{\mathrm{c}} - 2 \omega_{\mathrm{r}} )  \omega \nonumber \\ 
&& \quad + A \bigg( \frac{\omega_{\mathrm{z}}^2}{2} + i \alpha + \omega_{\mathrm{r}}^2 - \omega_{\mathrm{r}} ( \omega_{\mathrm{c}} + i \beta ) \bigg) - \epsilon B^{\ast} \bigg] \nonumber \\ 
+ \mathrm{e}^{- i \omega^{\ast} t} && \bigg[ B \omega^{\ast 2} + B ( i \beta + \omega_{\mathrm{c}} - 2 \omega_{\mathrm{r}} )  \omega^{\ast} \nonumber \\
&& \quad + B \bigg( \frac{\omega_{\mathrm{z}}^2}{2} + i \alpha + \omega_{\mathrm{r}}^2 - \omega_{\mathrm{r}} ( \omega_{\mathrm{c}} + i \beta ) \bigg) - \epsilon A^{\ast} \bigg] = 0. 
\end{eqnarray}}
It is straightforward to show that if $c \cdot f(x) + c^{\ast} \cdot g(x) = 0$, where $c$ has non-zero real and imaginary parts, then $f(x)$ and $g(x)$ must both be equal to zero.  Therefore 
{\setlength\arraycolsep{0pt}
\begin{eqnarray}
\label{eq:PenningaxializationvDEsolution2a}
A && \bigg[  \omega^2 -  \left( i \beta + \omega_{\mathrm{c}} - 2 \omega_{\mathrm{r}} \right)  \omega \nonumber \\
&& \quad + \left( \frac{\omega_{\mathrm{z}}^2}{2} + i \alpha + \omega_{\mathrm{r}}^2 - \omega_{\mathrm{r}} \left( \omega_{\mathrm{c}} + i \beta \right) \right) \bigg] - \epsilon B^{\ast} = 0\\ 
B && \bigg[ \omega^{\ast 2} + \left( i \beta + \omega_{\mathrm{c}} - 2 \omega_{\mathrm{r}} \right)  \omega^{\ast} \nonumber \\
\label{eq:PenningaxializationvDEsolution2b}
&& \quad + \left( \frac{\omega_{\mathrm{z}}^2}{2} + i \alpha + \omega_{\mathrm{r}}^2 - \omega_{\mathrm{r}} \left( \omega_{\mathrm{c}} + i \beta \right) \right) \bigg] - \epsilon A^{\ast} = 0.
\end{eqnarray}}
It was shown in section~\ref{sec:axialization} that in order to be in a frame in which the axializing quadrupolar drive is static it is necessary to have $\omega_{\mathrm{r}} = \omega_{\mathrm{a}} / 2$.  $\Delta$ is therefore defined to be half the detuning of the axialization drive from the cyclotron frequency, so that $\omega_{\mathrm{a}} = \omega_{\mathrm{c}} + 2 \Delta$.  Thus the frame rotation frequency is $\omega_{\mathrm{r}} = \omega_{\mathrm{c}} / 2 + \Delta$.  In addition to this the variable $\delta$ is defined to be the (complex) shift in the frequencies from $\omega_1$ due to the presence of laser cooling, the axializing drive and the change to the frame rotation frequency $\omega_r$, so that $\omega = \omega_1 + \delta$.  The equations for $\omega$ can therefore be rewritten as
{\setlength\arraycolsep{0pt}
\begin{eqnarray}
A && \bigg[ 2 \delta \omega_1 - i \beta \omega_1 + 2 \Delta \omega_1 \nonumber \\
\label{eq:PenningaxializationvDEsolution3a}
&& \quad \phantom{{}^{\ast \ast}} + i \frac{2 \alpha - \beta \omega_{\mathrm{c}}}{2} + \left( \Delta + \delta \right)^2 - i \beta \left( \Delta + \delta \right) \bigg] - \epsilon B^{\ast} = 0 \\
B && \bigg[ 2 \delta^{\ast} \omega_1 + i \beta \omega_1 - 2 \Delta \omega_1 \nonumber \\
\label{eq:PenningaxializationvDEsolution3b}
&& \quad + i \frac{2 \alpha - \beta \omega_{\mathrm{c}}}{2} + \left( \Delta - \delta^{\ast} \right)^2 - i \beta \left( \Delta - \delta^{\ast} \right) \bigg] - \epsilon A^{\ast} = 0.
\end{eqnarray}}
The $\left( \Delta \pm \delta \right)^2$ and $\beta \left( \Delta \pm \delta \right)$ terms are small relative to those containing $\omega_1$.  Ignoring these, and taking the complex conjugate of the latter equation, the pair of equations becomes
\begin{eqnarray}
\label{eq:PenningaxializationvDEsolution4a}
\; A & \bigg[ \left( 2 \delta \omega_1 - i \beta \omega_1 \right) + \left( 2 \Delta \omega_1 + i \frac{2 \alpha - \beta \omega_{\mathrm{c}}}{2}\right) \bigg] - \epsilon B^{\ast} & = 0\\
\label{eq:PenningaxializationvDEsolution4b}
B^{\ast} & \bigg[ \left( 2 \delta \omega_1 - i \beta \omega_1 \right) - \left( 2 \Delta \omega_1 + i \frac{2 \alpha - \beta \omega_{\mathrm{c}}}{2} \right) \bigg] - \epsilon^{\ast} A & = 0.
\end{eqnarray}			
Solving~(\ref{eq:PenningaxializationvDEsolution4b}) for $A$ and substituting into~(\ref{eq:PenningaxializationvDEsolution4a}) gives
\begin{equation}
\label{eq:PenningaxializationvDEsolution5}
B^{\ast} \bigg[ \left( 2 \delta \omega_1 - i \beta \omega_1 \right)^2 - \left( 2 \Delta \omega_1 + i \frac{2 \alpha - \beta \omega_{\mathrm{c}}}{2} \right)^2 - \left| \epsilon \right|^2 \bigg] = 0. 
\end{equation}
This can be simplified by defining $M = \left(2 \alpha - \beta \omega_{\mathrm{c}} \right) / 2 \omega_1$, a term related to the laser cooling parameters $\alpha$ and $\beta$.  In order to find the real and imaginary parts of this expression the substitution $\delta = \delta_0 + i \gamma_0$ is also made, so that  
\begin{equation}
\label{eq:Penningaxializationdsolution}
B^{\ast} \omega_1^2 \bigg[ \left( 2 \delta_0 + 2 i \gamma_0 - i \beta \right)^2 - \left( 2 \Delta + i M \right)^2 - \frac{\left| \epsilon \right|^2}{\omega_1^2} \bigg] = 0. 
\end{equation}
All of the variables are now real.  Evaluating the squared terms and separating out real and imaginary parts gives
\begin{equation}
\label{eq:Penningaxializationdrealsolution}
4 \delta_0^2 - 4 \gamma_0^2 - \beta^2 + 4 \beta \gamma_0 - 4 \Delta^2 + M^2 - \frac{\left| \epsilon \right|^2}{\omega_1^2} = 0
\end{equation}
and
\begin{equation}
\label{eq:Penningaxializationdimagsolution}
8 \delta_0 \gamma_0 - 4 \delta_0 \beta - 4 \Delta M = 0.
\end{equation}
The latter equation gives the cooling rates
\begin{equation}
\label{eq:appendixPenningaxializationgammaM}
\gamma_0 = \frac{\beta}{2} + \frac{\Delta M}{2 \delta_0}.
\end{equation}
Substituting $\gamma_0$ into~(\ref{eq:Penningaxializationdrealsolution}) then yields
\begin{equation}
\label{eq:PenningaxializationdeltaM}
\delta_0^2 - \left( \Delta^2 - \frac{M^2}{4} + \frac{\left| \epsilon \right|^2}{4 \omega_1^2} \right) - \frac{\Delta^2 M^2}{4 \delta_0^2} = 0,
\end{equation}
which, if $N$ is defined to be equal to the term in brackets, leads to the quartic equation
\begin{equation}
\label{eq:PenningaxializationdeltaMN}			
\delta_0^4 - N \delta_0^2 - \frac{\Delta^2 M^2}{4} = 0. 
\end{equation}
This has four roots
\begin{equation}
\label{eq:appendixPenningaxializationdeltaroots}
\delta_0 = \pm \frac{1}{\sqrt{2}} \sqrt{N \pm \sqrt{N^2 + \Delta^2 M^2}}.
\end{equation}
Since $\delta_0$ is by definition real, the positive sign must be taken for the inner square root and so two of the solutions can be eliminated.  Indeed these additional complex roots are in fact duplicated solutions.  If we use them to determine $\gamma_0$ we obtain identical expressions for $\delta = \delta_0 + i \gamma_0$. 

\section{Calculation of motional amplitudes for a driven, laser-cooled ion in the presence of axialization}
\label{sec:appendixB}

A dipolar field oscillating at frequency $\omega_{\mathrm{d}}$ in the laboratory frame can be thought of as a pair of static dipolar fields rotating in opposite senses at the oscillation frequency, since
\begin{equation}
\label{eq:drivingforce1}
2 F \cos{\omega_{\mathrm{d}} t} = F \left( \mathrm{e}^{i \omega_{\mathrm{d}} t} + \mathrm{e}^{-i \omega_{\mathrm{d}} t} \right).  
\end{equation}
If the driving frequency is close to the magnetron frequency, so that $\omega_{\mathrm{d}} = \omega_{\mathrm{m}} +\delta^{\prime}$, then in a frame rotating at a frequency $\omega_{\mathrm{r}} = ( \omega_{\mathrm{c}} / 2 ) + \Delta $ the driving force is given by
\begin{equation}
\label{eq:drivingforce2}
F \left( \mathrm{e}^{i \left(-\omega_1 + \delta^{\prime} - \Delta \right) t} + \mathrm{e}^{i \left( - \omega_{\mathrm{c}} + \omega_1 - \delta^{\prime} - \Delta \right) t} \right).
\end{equation}
The first term here is nearly resonant with the frequency of the magnetron motion in this frame (the detuning in the rotating frame being $\delta^{\prime} - \Delta$, henceforth defined to be $\delta$).  The second term will in general be far from resonance and can be ignored.  $\omega$ is defined to be the frequency of the driving field in the rotating frame, so that $\omega = \omega_{\mathrm{d}} - (\omega_{\mathrm{c}} / 2) - \Delta = -\omega_1 + \delta$.

Adding in this driving force into the equation of motion in the rotating frame for an ion being laser cooled with axialization~(\ref{eq:PenningaxializationvDE}) leads to
{\setlength\arraycolsep{0pt}
\begin{eqnarray}
\label{eq:drivenPenningaxializationvDE}
\ddot{v} + ( \beta - i \omega_{\mathrm{c}} + 2 i \omega_{\mathrm{r}} ) \dot{v} - \bigg( \frac{\omega_{\mathrm{z}}^2}{2} + i \alpha + \omega_{\mathrm{r}}^2 && - \omega_{\mathrm{r}} \left( \omega_{\mathrm{c}} + i \beta \right) \bigg) v + \epsilon v^{\ast}  \nonumber\\
&& = F \mathrm{e}^{i \omega t}. 
\end{eqnarray}}
Steady state solutions corresponding to the situation once the system has reached equilibrium will have the form of elliptical motion with a frequency $\omega$, such that
\begin{equation}
\label{eq:PenningaxializationvDEtrialrepeatagain}
v = A \mathrm{e}^{i \omega t} + B \mathrm{e}^{-i \omega^{\ast} t}.
\end{equation}
Here $\omega$ is real and is the known applied frequency rather than a variable.  Proceeding in an identical manner to before leads to the equivalent of~(\ref{eq:PenningaxializationvDEsolution4a}) and~(\ref{eq:PenningaxializationvDEsolution4b}):
\begin{eqnarray}
\label{eq:drivenPenningaxializationvDEsolution1a}
\; A & \bigg[ -2 \delta \omega_1 + i \beta \omega_1 -  2 \Delta \omega_1 - i \alpha + i  \beta \frac{\omega_{\mathrm{c}}}{2} \bigg] + \epsilon B^{\ast} & = F,\\
B^{\ast} & \bigg[  -2 \delta \omega_1 + i \beta \omega_1  +  2 \Delta \omega_1 + i \alpha - i \beta \frac{\omega_{\mathrm{c}}}{2} \bigg] + \epsilon^{\ast} A & = 0.
\end{eqnarray}
The first of these equations is for the component rotating in the same sense as the driving force (in this case the direction of the magnetron motion), with the second equation describing the component of rotation in the opposite sense.  Bearing in mind that $\delta$ is now a defined constant detuning, the constant terms in square brackets can be grouped together as $C_{\mathrm{m}}$ and $C_{\mathrm{c}}$ respectively.  This gives simply
\begin{equation}
\label{eq:drivenPenningaxializationvDEsolution2a}
A C_{\mathrm{m}} + \epsilon B^{\ast} = F,
\end{equation}
\begin{equation}
\label{eq:drivenPenningaxializationvDEsolution2b}
B^{\ast} C_{\mathrm{c}} + \epsilon^{\ast} A = 0. 
\end{equation}
These equations are easily solved for $A$ and $B$, yielding
\begin{equation}
\label{eq:drivenPenningaxializationvDEsolution3a_appendix}
A = \frac{F C_{\mathrm{c}}}{C_{\mathrm{m}} C_{\mathrm{c}} - \left| \epsilon \right|^2},
\end{equation}
\begin{equation}
\label{eq:drivenPenningaxializationvDEsolution3b_appendix}
B = \frac{F \epsilon}{\left| \epsilon \right|^2 - C^{\ast}_{\mathrm{m}} C^{\ast}_{\mathrm{c}}}.
\end{equation}


\section*{References}
\bibliography{axialisation_theory_paper}
\bibliographystyle{unsrt}

\end{document}